\documentclass[twocolumns,reprint,prd,tightenlines,eqsecnum,floats,aps,amsmath,amssymb,nofootinbib,superscriptaddress,showpacs]{revtex4-1}   
\usepackage{amsmath,amsthm,amssymb,amsfonts}
\usepackage{MnSymbol}
\usepackage{colordvi}

\numberwithin{equation}{section}
\numberwithin{thr}{section}
\numberwithin{chr}{section}
\numberwithin{df}{section}

\newcommand{\scpr}[2]{\langle#1\, \vert \, #2 \rangle}
 
\newcommand{\norm}[1]{\left\lVert #1 \right\rVert}
\newcommand{\ket}[1]{\lvert\, #1\,\rangle}
\newcommand{\bra}[1]{\langle\, #1\,\rvert}

\newcommand{\gbra}[1]{(\,#1\,\rvert}

\newcommand{\comm}[2]{\left[\,#1\, ,\, #2\, \right]}

\newcommand{\GR}{{\rm GR}}
\newcommand{\SF}{{\rm SF}}

\DeclareMathOperator{\cyl}{Cyl}

\DeclareMathOperator{\sspan}{Span}
\begin{document}

\title{Loop quantum gravity coupled to a scalar field}

\author{Jerzy Lewandowski}\email{Jerzy.Lewandowski@fuw.edu.pl}
\affiliation{Faculty of Physics, Uniwersytet Warszawski,
ul. Pasteura 5, 02-093 Warszawa, Polska (Poland)}
\affiliation{Institute for Quantum Gravity, Department of Physics
Friedrich-Alexander-Universit\"at  Erlangen-N\"urnberg (FAU), 91058 Erlangen (Germany)}

\author{Hanno Sahlmann}
\email{hanno.sahlmann@gravity.fau.de}
\affiliation{Institute for Quantum Gravity, Department of Physics
Friedrich-Alexander-Universit\"at  Erlangen-N\"urnberg (FAU), 91058 Erlangen (Germany)}

\begin{abstract} 
We reconsider the Rovelli-Smolin model of gravity coupled to the Klein-Gordon time field with an eye towards capturing the degrees of freedom of the scalar field lost in the framework in which
time is deparametrized by the scalar field. Several new results for loop quantum gravity are obtained: 

(i) a Hilbert space for the gravity-matter system and a non-standard representation of the scalar field thereon is constructed, 
(ii) a new operator for the scalar constraint of the coupled system is defined and investigated,
(iii) methods for solving the constraint are developed. 

Commutators of the new constraint \emph{do not} vanish, but seem to reproduce a part of the Dirac algebra. This, however, poses problems for finding solutions. Hence the states we consider -- and perhaps the whole setup -- still needs some improvement. 

As a side result we describe a representation of the gravitational degrees of freedom in which the flux is diagonal. This representation bears a strong resemblance to the BF vacuum of Dittrich and Geiller. 
\end{abstract} 
      
\maketitle

\newpage

\tableofcontents

\section{Introduction} 
\label{se_intro}
Considering matter coupled to gravity is obviously very relevant for physical reasons. But there is an added benefit in canonical quantizations of gravity. Components of matter fields can act as preferred coordinates and thus help with the so called \emph{problem of time}. 
Rovelli and Smolin \cite{gr-qc/9308002} proposed the use of a scalar field as time in loop quantum gravity (LQG). Recently, it has been shown \cite{arXiv:1009.2445} that a full solution to the dynamics of LQG can be achieved, using a scalar field in this way. Dust  (as first pointed out in \cite{gr-qc/9409001}) has also been used for deparametrization in LQG \cite{arXiv:0711.0119,arXiv:1108.1145}.

In the present work, we come back to the coupling of gravity to a scalar field. But unlike in \cite{gr-qc/9308002, arXiv:1009.2445}, we do not want to trade the scalar field for a preferred time already at the classical level, and quantize a system with a true Hamiltonian. Rather, the idea is to treat the coupled system as a constrained system in the usual way, and only look for a connection to the deparametrized system \emph{after} quantization and -- possibly -- solution of the constraints.

We employ a diffeomorphism invariant representation based on a Hilbert space $\mathcal{H}_\SF$ for the scalar field $\phi$ in which 
$\widehat{\phi}$ is diagonal
\begin{equation*}
\label{ourrep}
 \widehat{\phi}(x)\ \ket{\varphi}\ =\ \varphi(x) \ket{\varphi},
 \end{equation*}
while its momentum is only well defined in an exponentiated version. A representation with similar properties has been used in the context of quantum cosmology before, see \cite{arXiv:0906.2798,arXiv:1002.0138}. Also, this representation is closely related to one for gravity developed by Campiglia and Varadarajan \cite{arXiv:1306.6126,arXiv:1311.6117,arXiv:1406.0579}.\footnote{In that representation, the flux operators have eigenvalues that are a sum of a background part, and the usual values from the Ashtekar-Lewandowski representation, and there is an operator quantizing the functional $F[A]=\exp \int E A\, \text{d}^3x$. This operator changes the flux background. In the present case, this role is played by the exponentiated momentum $\exp(\pi(f))$ of the scalar, and the usual Ashtekar-Lewandowski sector is absent. In the terminology of Campiglia and Varadarajan, the representation \eqref{ourrep} is the \emph{pure background case}.} The representation is inequivalent to the standard representation \cite{Ashtekar:2002vh}, but there is an interesting relation between the two, which we will elucidate. 

The new representation allows us to quantize all ingredients for the scalar constraint of the combined system, including the piece 
\begin{equation*}
\label{eq_qop}
\widehat{\sqrt{{\phi}_{,a}{\phi}_{,b}{E}^a_{i}{E}^b_{i}}}(x).
\end{equation*}
Geometrically, it contains the area element of the level surfaces of the scalar field. The corresponding operator has a simple action on the Hilbert space of the combined system, and slightly generalizes the $\widehat{Q}$\emph{-operator} of \cite{gr-qc/0005117}.  

In this way, we succeed in obtaining a constraint operator $\widehat{C}(x)$ acting in a certain subset of the Hilbert space $\mathcal{H}_\SF\otimes\mathcal{H}_\GR$, or rather, on some subspace of its dual. We find that 
\begin{widetext}
\begin{equation*}
\label{eq_dirac}
[\widehat{C}(M),\widehat{C}(N)]\; \eta(\bra{\varphi}\otimes\bra{\gamma,j,\iota}) 
=8\pi \beta \ell_\text{P}^2\left(\sum_{e\in\gamma}\sqrt{j_e(j_e+1)}\int_e  {\rm sgn} (d\varphi) \left(NdM-MdN\right)\right) \;\eta(\bra{\varphi}\otimes\bra{\gamma,j,\iota}). 
\end{equation*}
\end{widetext}
Here, $\eta$ is a partial group averaging over diffeomorphisms. It appears that the term in bracket could be interpreted as a quantization of the expected contribution 
\begin{equation*}
\int  S^a(x)\;  \pi \partial_a\phi (x)\;  \text{d}^3 x, \qquad S^a = \frac{E_i^a E_j^b\delta^{ij}}{|\det E |} \left(N\partial_bM-M\partial_bN\right) 
\end{equation*}
of the scalar field to the right hand side. While this is certainly intriguing, it also presents an obstruction to finding solutions of the scalar constraint, as it turns out that the states we devise in the present article are not annihilated by the right hand side  of \eqref{eq_dirac}. Hence the states we consider -- and perhaps the whole setup -- still needs some improvement. 

We do, however find non-trivial solution spaces for some truncated versions of the Hamilton constraint. They lie in the tensor product of the dual to (a dense subspace of) $\mathcal{H}_\SF$ and the kinematic space of the gravitational sector. 

As a side result of the discussion of various diffeomorphism invariant representations for the scalar field, we also describe a representation of the gravitational degrees of freedom in which the flux is diagonal. This representation bears a strong resemblance to the BF vacuum representation of Dittrich and Geiller \cite{arXiv:1401.6441,arXiv:1412.3752}. 

The work is organized as follows: Discussion of scalar field representation and BF-representation is found in sec.\ \ref{se_kin}. Sec.\ \ref{se_comp} contains the quantization of various components of the constraint, including  that of \eqref{eq_qop}.

\section{Classical theory} 
\label{Sec:Classical Theory}
In this article, we will consider 4d Einstein gravity coupled to a scalar field, given by the action
\begin{equation*}S[\phi,e,\omega]= S_\text{GR}+S_\text{Holst}+S_\text{Scalar}\end{equation*}
with 
\begin{align*}
S_\text{GR}&=\frac{1}{32\pi G}\int \epsilon_{IJKL} e^I\wedge e^J\wedge F^{KL}(\omega)\\
S_\text{Holst}&=-\frac{1}{16\beta G}\int e^I\wedge e^J\wedge F_{KL}(\omega)\\
S_\text{Scalar}&=\frac{1}{2\lambda}\int \sqrt{g(e)}\left(g^{\mu\nu}(e)(\partial_\mu\phi)(\partial_\nu\phi)-u(\phi)\right)
\end{align*}
The canonical analysis of this action, and a partial gauge fixing (time gauge) leads to a phase space coordinatized by fields
\begin{equation*}
(\phi(x),\pi(x),  A_a^i(x),E^b_j(x))
\end{equation*}
taking values on a spatial slice $\Sigma$ of space-time. For a detailed derivation see for example \cite{gr-qc/0404018}. Indices $a,b,\ldots$ are spatial, whereas $i,j,\ldots$ refer to su(2), the algebra of the gauge group after partial gauge fixing. For 4d space-time, the manifold $\Sigma$ is 3-dimensional. Our results for the scalar field sector generalize to other space-time dimensions, whereas the treatment of the gravitational field is special to four dimensions.  

$\phi$  is the scalar field, and $\pi$ its canonically conjugate momentum. The scalar field is a function
\begin{equation*} \phi : \Sigma\rightarrow \mathbb{R}\end{equation*}     
while the momentum $\pi$ is a weight-1 density. Thus it defines 
a pseudo  $3$-form
\begin{equation*}  \pi dx^1\wedge dx^2\wedge dx^3\end{equation*} 
that changes the sign whenever the coordinate transformation has a negative Jacobian. In the space 
\begin{equation*}\Gamma_\SF\ = \{(\phi,\pi) \ :\ {\rm as\ above}\}\end{equation*}
we consider the canonical Poisson bracket 
\begin{equation*}
 \{F(\phi,\pi),\ G(\phi,\pi)\}\ =\ \int_\Sigma \text{d}^3x\, \frac{\delta F}{\delta \phi(x)} \frac{G}{\delta \pi(x)}
\ -\ \frac{\delta G}{\delta \phi(x)} \frac{\delta F}{\delta \pi(x)}.
\end{equation*}

The classical phase space $\Gamma_\GR$  for the gravitational field consists of the su(2) valued                              
1-form field
\begin{equation*} 
A(x)\ =\  A_a^i(x)\; \tau_i \otimes dx^a\ . 
\end{equation*} 
and the canonically conjugate momentum vector-density
\begin{equation*} 
E(x)\ =\  E^a_i(x)\;\tau^{i*}\otimes \partial_a\ .
\end{equation*}
The usual choice of the basis $\tau_1,\tau_2,\tau_3$ is such that
\begin{equation*} [\tau_i,\tau_j]=\sum_k \epsilon_{ijk}\tau_{k}\end{equation*}
and $\tau^{*i}$ is the dual basis in su(2)$^*$. 
The Poisson bracket between two functionals $F[A,E],\ G[A,E]$ is
\begin{equation*} \{F,G\} = 8\pi G \beta
\int_\Sigma \text{d}^3x \frac{\delta F}{\delta A^i_a(x)} \frac{\delta G}{\delta E^a_i(x)}
- \frac{\delta G}{\delta A^i_a(x)} \frac{\delta F}{\delta E^a_i(x)} .
\end{equation*}
The phase space $\Gamma=\Gamma_\SF\times\Gamma_\GR$  is not yet physical, however. Rather, the physical phase space is induced by constraints on $\Gamma$. The main concern of the present work is the implementation in the quantum theory of the scalar constraint
\begin{equation}\label{scalar}
\begin{split}
 C(x)\ =\ &\frac{\pi^2(x) + \phi_{,a}(x)\phi_{,b}(x)E^a_{i}(x)E^b_{i}(x)}{2\sqrt{|\det E(x)|}}\\
  &\qquad + u(\phi(x)){\sqrt{|\det E(x)|}} + C_\GR(A,E)(x) 
 \end{split}
\end{equation}
where 
\begin{equation}\label{vacuum_scalar}
C_\GR = \frac{1}{16\pi G}  
\frac{E^a_iE_j^b}{\sqrt{|\det E|}}\left(\epsilon^{ijk}F_{abk}+2(\sigma-\beta^2)K_{[a}^iK_{b]}^j \right)
\end{equation}
is the scalar constraint of vacuum gravity. $F$ is the curvature of $A$ and $K$ is the extrinsic curvature of $\Sigma$, which is a function of $A$ and $E$.  For the Lorentzian gravity $\sigma=-1$. The Euclidean model of gravity is
defined by $\sigma=1$.
\section{Kinematic quantization}
\label{se_kin}
In the present section, we will quantize the kinematic phase space $\Gamma=\Gamma_\SF\times\Gamma_\GR$, resulting in a Hilbert space ${\cal H}\ =\  {\cal H}_\SF\otimes{\cal H}_\GR$. The gravitational sector is quantized exactly as usual in LQG. The quantization of the scalar field sector is non-standard. 
\subsection{Diffeomorphism invariant position and  momentum representations for the scalar field}
\subsubsection{The position representation}
We introduce now a representation for scalar field that will be used through out this 
paper.  A similar representation has been  defined  for a tensor field in \cite{arXiv:0904.0184}, and its finite dimensional version has been applied in quantum cosmology and other contexts (see ex.\ \cite{arXiv:0906.2798, arXiv:1002.0138}). Moreover, this representation is precisely the scalar field version of the \emph{pure background representation} \cite{arXiv:1306.6126,arXiv:1311.6117,arXiv:1406.0579} (see also \cite{Koslowski:2007kh,Sahlmann:2010hn,arXiv:1312.3657}). 
 
A quantum state $\ket{\varphi}$ is defined by any function
\begin{equation*} \varphi\ :\ \Sigma\ \rightarrow\ \mathbb{R} \end{equation*}
where the differentiability class $C^k$ is to be fixed depending on applications. 
The Hilbert scalar product is defined to be
\begin{equation*} \scpr{\varphi}{\varphi}\ =\  1,\quad \text{ and } \quad \scpr{\varphi}{\varphi'}\ =\ 0 \ \ {\rm whenever}\  \  \varphi\not=\varphi' ,\end{equation*}
and the Hilbert space ${\cal H}_\SF$ is spanned by the quantum states defined above.
The quantum scalar field operator $\widehat{\phi}(x)$ is defined to act as
\begin{equation*}  \widehat{\phi}(x)\ \ket{\varphi}\ =\ \varphi(x) \ket{\varphi}\end{equation*}
while the best we can do for representing the quantum momentum operator is
to define
\begin{equation} \label{hat{pi}} {\exp\left(-\frac{i}{\hbar}\int \text{d}^3x\, f(x)  \widehat{\pi}(x) \right)\ } \ket{\varphi}\ =\ \ket{\varphi+f} .
\end{equation} 

${\cal H}_\SF$ can be thought of as spanned by the functions $ \ket{\varphi}:\Gamma_\SF\rightarrow\mathbb{R}$
of momenta
\begin{equation}\label{|varphi>} \ket{\varphi}[\pi]\ =\  \exp\left(-\frac{i}{\hbar}\int \text{d}^3x\, \varphi(x) {\pi}(x) \right), \end{equation}
endowed with the functional integral
\begin{align*}
 \int \prod_{x}\text{d}\pi(x)  \exp\left[-\frac{i}{\hbar}\int \text{d}^3x f(x)  {\pi}(x) \right]\ &:=\ 0,\\
 \int \prod_{x}\text{d}\pi(x) \; 1\ & :=\ 1\ .
 \end{align*}

That is why $\widehat{\pi}(x)$ is not defined in this Hilbert space itself. 
On the other hand, every expression $\phi_{,a\ldots b}\equiv \partial_a\ldots\partial_b\phi$ is represented by the well defined operator
\begin{equation*} \widehat{\phi}_{,a...b}(x)\ \ket{\varphi}\ =\ {\varphi}_{,a...b}(x) \ket{\varphi}\ .\end{equation*}
The dense set of finite linear combinations
\begin{equation}
\label{cyl}
 \cyl_\SF = \sspan\{ \ket{\varphi}\; |\; \varphi \in C^k(\Sigma) \}
\end{equation}
is called the set of cylindrical functions of the scalar field. A set of generalized states is given by the linear functionals $\cyl_\SF^{*}$ over $\cyl_\SF$ which can be explicitly written as 
\begin{equation}
\label{generalized}
\cyl_\SF^*=\{   \gbra{\psi},\; \psi: C^k(\Sigma)\rightarrow \mathbb{C}\}, \quad 
\gbra{\psi}=\sum_{\varphi} \psi[\varphi] \bra{\varphi}. 
 \end{equation}   
 
The diffeomorphisms  of $\Sigma$ act naturally in ${\cal H}_\SF$. For $f\in$Diff we have
\begin{equation*}
(U_f \ket{\varphi})[\pi]\ :=\ \ket{\varphi}[f^*\pi]\ =\ \ket{\varphi\circ f^{-1}}[\pi].
\end{equation*}   
Or, without thinking of the vectors as functions,
\begin{equation*}
U_f \ket{\varphi}\ =\ \ket{\varphi\circ f^{-1}}.
\end{equation*} 
This is a unitary action. By the duality
\begin{align*}
U_f^*\bra{\varphi} \ =\  \bra{\varphi \circ f}
\end{align*}   
and finally
\begin{align*}
U_f^*\gbra{\psi}\ =\ \gbra{\psi'},\qquad \psi'[\phi]\ =\ \psi[\phi\circ f^{-1}]
\end{align*}
\subsubsection{The momentum representation}
In analogy to the position representation of the previous section, we can also define a momentum representation. The \emph{polymer representation} \cite{Ashtekar:2002vh}, which is the standard representation for the scalar field in LQG, will emerge as a restriction of the momentum representation. 

Define a quantum state $\ket{p}'$  by \emph{any} weight 1 density
\begin{align*} p\ :\ \Sigma\ \rightarrow\ \mathbb{R} .
\end{align*}
The Hilbert scalar product is defined to be
\begin{equation*} \scpr{p}{p}'\ =\  1,\quad \text{ and } \quad \scpr{p}{p'}'\ =\ 0 \ \ {\rm whenever}\  \ 
 p\not=p' ,\end{equation*}
and the Hilbert space ${\cal H}'_\SF$ is spanned by the quantum states defined above.
The quantum scalar field  momentum  operator $\widehat{\pi}(x)$ is defined to act as
\begin{equation*} 
 \widehat{\pi}(x)\ \ket{p}'\ =\ p(x) \ket{p}'
\end{equation*}
while the best we can do for representing the quantum scalar field operator is
to define
\begin{equation}
\label{hatphi} 
\exp\left(\frac{i}{\hbar}\int \text{d}^3x\, p'(x)  \widehat{\phi}(x) \right)\  \ket{p}'\ =\ 
\ket{p+p'}' .
\end{equation} 

${\cal H}'_\SF$ can be thought of as spanned by the functions 
$ \ket{p}':\Gamma_\SF\rightarrow\mathbb{C}$
of scalar field
\begin{equation}
\label{|p>}
 \ket{p}'[\phi]\ =\  \exp\left(\frac{i}{\hbar}\int \text{d}^3x\, p(x) {\phi}(x) \right),\end{equation}
endowed with the integral
\begin{align*} 
\int \prod_{x}\text{d}\phi(x)  \exp\left[-\frac{i}{\hbar}\int \text{d}^3x\; p(x)  {\phi}(x) \right]\ &:=\ 0,\\
 \int \prod_{x}\text{d}\phi(x)\;  1\ &:=\ 1\ .
\end{align*}
That is why $\widehat{\phi}(x)$ is not defined in this Hilbert space itself.  

The space of the weight 1 densities contains differentiable densities,  however it also contains 
distributional densities, for example
\begin{equation}
\label{pdelta}
p(x)\ =\ \sum_{i=1}^k p(x_i) \delta(x,x_i) .
\end{equation}
If we restrict the states  to densities $p$ defined by \eqref{pdelta} then we obtain the known polymer 
representation ${\cal H}_{Pol}$ \cite{Ashtekar:2002vh}.  

We note that there are other similar distributional subspaces, for example the space of densities of the form 
\begin{equation*}
p(x)\ =\ \sum_{i=1}^k X^a(x_i)\partial_a \delta(x,x_i). 
\end{equation*}
These have not been applied in LQG thus far.  
 
The diffeomorphisms  of $\Sigma$ act naturally in ${\cal H}'_\SF$. For $f\in$Diff we have
\begin{equation*}
(U_f \ket{p}')[\phi]\ :=\ \ket{p}'[f^*\phi]\ =\ \ket{f^{-1*}p}'[\phi].
\end{equation*}   
Shortly,
\begin{equation*}
U_f \ket{p}'\ =\ \ket{f^{-1*}p}'.
\end{equation*} 
This is a unitary action. By the duality
\begin{align*}
U_f^*\bra{p} \ =\  \bra{f^*p} .
\end{align*}   
\subsubsection{The non-unitary duality between the position and momentum representations}
The position and momentum representations are dual to each other. Each state $\ket{p}'\in{\cal H}'_\SF$ 
can be mapped into a state dual to $\cyl_\SF$,
\begin{equation*}
\ket{p}'\ \mapsto\ \gbra{\Psi_p}\in \cyl_\SF^*, 
\end{equation*}
defined to be
\begin{equation}\label{duality}
\gbra{\Psi_p}(\ket{\varphi})\ =\  e^{\frac{i}{\hbar}\int \text{d}^3x p(x)\varphi(x)} .
\end{equation}     
In other words, via this duality, $\bra{p}$ is naturally mapped into  a generalized function, an example of an element in $\cyl_\SF^*$, see \eqref{generalized}, 
\begin{equation}\label{ptovarphi}
\gbra{\Psi_p}\ =\ \sum_{\varphi}  \Psi_p[\varphi] \bra{\varphi}, \quad \text{with} \quad  \Psi_p[\varphi]=  e^{\frac{i}{\hbar}\int \text{d}^3x p(x)\varphi(x)} .
\end{equation}

\subsection{Kinematic quantization of the gravitational field} 
We consider the kinematic quantization of gravity in the framework of loop quantum gravity. The LQG quantum states are cylindrical functions of $A$,
\begin{equation*}  \Psi[A]\ =\ \psi(h_{e_1}[A],...,h_{e_n}[A]) \end{equation*}
which can be expressed by parallel transports
\begin{equation*} h_e[A]\ =\ \mathcal{P}\exp \left(-\int_e A\right) 
\end{equation*} 
where $e_1,...,e_N$ are finite curves -- we will also refer to them as  \emph{edges} -- in $\Sigma$ which form an embedded 
graph $\gamma=\{ e_1,...,e_n \}$ (i.e., no (self-) intersections except for possibly at the ends).     

An important class of functions is  spanned by the polynomials 
\begin{equation}\label{sn}
 \ket{\gamma,j,\iota}[A]\ =\ \iota_{A_1...A_n}^{B_1...B_n} \; \rho^{(j_1)}{}^{A_1}_{B_1}(h_{e_1}[A])\ldots \rho^{(j_n)}{}^{A_n}_{B_n}(h_{e_n}[A])
 \end{equation}
 where  $\rho^{(j_1)},...,\rho^{(j_n)}$ are the representations of SU(2) corresponding to the spins $j_i=\frac{1}{2},1,...$, 
 and  $\iota$ is an arbitrary tensor. We will denote it 
 $$ {\rm Cyl}_\GR\ :=\ {\rm Span}(\ket{\gamma,j,\iota})$$ 
 throughout this paper, and use it as the domain for our operators and distributions.  
 
 Given $\gamma$ and $\rho$, a suitable choice of $\iota$  makes the function $\ket{\gamma,j,\iota}$ invariant with respect to the SU(2) gauge transformations $A\mapsto g^{-1}Ag+g^{-1}dg$, $g:\Sigma\rightarrow $SU(2), but at this point we are not assuming that property.

To calculate the scalar product between two cylindrical functions $\Psi$ and $\Psi'$
defined by using graphs $\gamma$ and  $\gamma'$, respectively, we find a refined 
graph $\gamma''=\{e''_1,...,e''_{n''}\}$, such that  both the functions can be written as
\begin{align*}
\Psi[A] &= \psi(h_{e''_1}[A],...,h_{e''_{n''}}[A]),\\ 
\Psi'[A] &= \psi'(h_{e''_1}[A],...,h_{e''_{n''}}[A]).
\end{align*}
The existence of a common refined graph $\gamma''$ for every pair of graphs $\gamma$ and $\gamma'$ is ensured
by assuming the (semi)analyticity of $\Sigma$ and the edges of the graphs \cite{Ashtekar:1994wa,Lewandowski:2005jk}.   
The scalar product is defined to be
\begin{equation}
\label{scpr} \scpr{\Psi}{\Psi'}\ =\ \int dg_1...dg_{n''}\overline{\psi}(g_1,...,g_{n''}){\psi}'(g_1,...,g_{n''}),
\end{equation}
where the integral is independent of the choice of refined graph $\gamma"$ due to the properties of the Haar measure \cite{Ashtekar:1994mh}.
 
The cylindrical functions span the Hilbert space 
\begin{equation*}
{\cal H}_\GR=\overline{\cyl_\GR}^{\norm{\cdot}}
\end{equation*}
where the norm is that from the scalar product \eqref{scpr}. 
  
Every cylindrical function $\Psi$ is also a multiplication operator
\begin{equation*}
(\Psi(\widehat{A})\Psi')[A]\ =\ \Psi[A]\Psi'[A].
\end{equation*}  
A connection operator $\widehat{A}$ by itself is not defined. The field $E$ is naturally quantized as 
\begin{equation*}
\widehat{E}^a_i\Psi[A]\ =\ \frac{\hbar}{i}\{\Psi[A],E^a_i(x)\} \ =\ \frac{8\pi\beta \ell^2_P}{i}\frac{\delta}{\delta A^i_a}\Psi[A]. 
\end{equation*}
It turns into well defined operators in ${\cal H}_\GR$ after smearing  either along a 2-surface $S\subset \Sigma$
$$
\widehat{E}_{S,f}:=\frac{1}{2}
\int_S f^i\widehat{E}^a_i\epsilon_{abc}dx^b\wedge dx^c\ \ \ \ \  f:S\rightarrow {\rm su}(2)$$
where $f$ may involve parallel transports \cite{Rovelli:1989za,Thiemann:2000bv}: 
$$ f(x)\ =\ \tilde{f}(x)\left(h_{p{x_0x}}\xi h_{p{xx_0}}\right)^i     $$
where $S\ni x\mapsto p_{xx_0}$ assigns to each point $x\in S$ a path $p_{xx_0}$ connecting a fixed point $x_0$ to $x$,
$\xi\in$su(2), and $\tilde{f}:S\rightarrow \mathbb{R}$.     

The diffeomorphisms  Diff of $\Sigma$ act naturally in ${\cal H}_{\GR}$. For $f\in$ Diff we have
\begin{equation*}
\begin{split}
(U_f \Psi)[A]\ :=&\ \Psi[f^*A]\ =\ \psi(h_{e_1}[f^*A], ...,h_{e_n}[f^*A])\\ 
=&\ \psi(h_{f(e_1)}[A], ...,h_{f(e_n)}[A]).
\end{split}
\end{equation*}   
Therefore
\begin{equation*}
U_f \ket{\gamma,j,\iota}\ =\ \ket{f(\gamma),j\circ f^{-1}, \iota\circ f^{-1}}.
\end{equation*} 
This is a unitary action. By the duality
\begin{align*}
U_f^*\bra{\gamma,j,\iota} \ =\  \bra{f^{-1}(\gamma),j \circ f,\iota\circ f} .
\end{align*}   
  
\subsection{The Ashtekar-Barbero connection position representation and the BF vacuum}
A quantum position representation 
is defined by states $\ket{a}$ labelled by
su(2) valued 1-forms 
$$ a(x)=a^i_b(x)\tau_idx^b.$$
The states are normalized and orthogonal to each other
$$\scpr{a}{a'}\ =\ 1,\ \ \ \ \ \ \ \ \ \scpr{a}{a'}\ =\ 0, \ {\rm for\ } a\not=a' .$$  
The representation is
$$\widehat{A}^i_b(x)\ket{a}\ =\ a^i_b(x)\ket{a}$$ 
and
\begin{equation}
\label{eE}
\exp\left(\widehat{E}(\omega)/8\pi i\beta \ell_{\rm P}^2\right)\,\ket{a}=\ \ket{a+\omega}, \text{ where }\widehat{E}(\omega)=\int \widehat{E}^a_i\omega^i_a\,\text{d}^3x
\end{equation}
for every smearing 1-form $\omega$. This representation is related to the pure background representation of \cite{arXiv:1306.6126,arXiv:1311.6117,arXiv:1406.0579}, but with the roles of position and momentum reversed. 
One can think of this representation as given by the scalar product
$$ \int \prod_x \text{d}E(x) 1\ =\ 1, \ \ \ \ \ \ \  \int \prod_x \text{d}E(x)  \exp(E(\omega)/8\pi i\beta \ell_{\rm P}^2) =\ 0. $$
and 
\begin{equation*}
\begin{split}
\widehat{A}^i_a(x)\ &=\ 
{8\pi i\beta \ell_{\rm P}^2}
\frac{\delta}{\delta E^a_i(x)},\\
\exp(\widehat{E}(\omega)/8\pi i\beta \ell_{\rm P}^2)\ &=\ \exp({E}(\omega)/8\pi i\beta \ell_{\rm P}^2).
\end{split}
\end{equation*}
Notice that in this representation, there is an operator
corresponding to the classical 
$$ F\ =\ dA + \frac{1}{2}[A,A] \ =\ \frac{1}{2} F^i_{ab}dx^a\wedge \tau_i\otimes dx^b .$$ 
Namely, 
$$ \widehat{F}^i_{ab}\ket{a}\ =\ f^i_{ab}\ket{a} , \ \ \ \ \ f\ =\ da + \frac{1}{2}[a,a] . $$
Therefore,  this representation deserves the name   BF representation,
because it contains states  such that
$$ \widehat{F}^i_{ab}\ket{a_0}\ =\ 0.$$      

Each state $\ket{a}$ can be identified with a function
\begin{equation*}
 \ket{a}: E\ \mapsto\ \exp\left(E(a)/8\pi i\beta \ell_{\rm P}^2 \right)
\end{equation*}
The 1-forms $a$ labeling the states can well be distributional connections supported 
at the sites of 2-complexes, rather than smooth 1-forms. Specifically, such a connection
is defined by the following data: 
\begin{itemize}
\item a 2-complex $\Delta$
\item orientation of its 2-faces
\item an su(2) valued function $a:\Delta\rightarrow$su(2). 
\end{itemize} 
This data is a generalization of a connection because it defines a parallel transport
along each curve $p$ in $\Sigma$:
$$ p\ \mapsto\ h_p(a)\in SU(2)$$ 
where  
$$ h_p(a)\ =\      e^{\mp a(x_k)}\cdot \ldots \cdot e^{\mp a(x_1)}, $$
$x_1,...,x_k$ are the points at which the curve $p$ crosses faces of the 2-complex,
the sign depends on the relative orientation of $p$ and the face of $\Delta$ at the 
intersection point (we may have to assume, that only intersections not contained
in the edges contribute), and the formula is naturally generalized to curves
which end or begin at a face of $\Delta$. 
The data itemized above defines also a function of $E$,
$$ \exp\left(E(a)/8\pi i\beta \ell_{\rm P}^2\right), \qquad \text{ with } E(a)=\frac{1}{2}\int_\Delta  a^iE_i^{a}\epsilon_{abc}\text{d}x^b\wedge \text{d}x^c $$ 
and a corresponding quantum operator
$$ \exp\left(\widehat{E}(a)/8\pi i\beta \ell_{\rm P}^2\right)\,\ket{a'}\ =\ \ket{a+a'}.$$ 

It appears to us that with these choices, we reproduce many aspects of the \emph{BF vacuum representation} of Dittrich and Geiller \cite{arXiv:1401.6441,arXiv:1412.3752}
\section{Quantizing the components  of the scalar constraint}
\label{se_comp}
Because of the scalar constraint \eqref{scalar}, on the constraint surface we have
\begin{equation}\label{C}
\pi = \sqrt{ - \phi_{,a}\phi_{,b}E^a_{i}E^b_{i} -2u(\phi)|\det E| -2 \sqrt{|\det E|}C_\GR(A,E) } .
\end{equation}
in the classical theory. We will now have to make sense of this equation in the quantum theory. To this end, we will presently 
consider the quantization of the components of this equation, namely the quantization of 
$\sqrt{ - \phi_{,a}\phi_{,b}E^a_{i}E^b_{i}}(x)$ (section \ref{ppEE}), 
$\sqrt{ - \phi_{,a}\phi_{,b}E^a_{i}E^b_{i} -2u(\phi)}(x)$ (section \ref{A+B}), $\pi(x)$ (section \ref{pi_dual}), and 
$C_\GR(A,E) (x)$ (section \ref{Cgr}). We work in the Hilbert space $\mathcal{H}_\SF\otimes\mathcal{H}_\GR$ and sometimes in its dual. 
\subsection{Quantization of $\sqrt{\partial \phi \partial \phi EE  }$}
\label{ppEE}
We now consider the following classical expression
\begin{equation*} \int \text{d}^3 x f(x) \sqrt{{\phi}_{,a}(x){\phi}_{,b}(x){E}^a_{i}(x){E}^b_{i}(x)},\end{equation*}
where $f$ is an arbitrary smearing function.
The key observation about this very expression can be made already in the classical level.   
Given a point 
\begin{equation*}
(\phi,\pi,A,E)\ \in\ \Gamma_\SF\times\Gamma_\GR
\end{equation*}
 consider  a neighborhood  in $\Sigma$ in which 
\begin{equation*} d\phi \ \not=\ 0. \end{equation*} 
Let us choose coordinates $(x^1, x^2, x^3)=(y^1,y^2,\phi)$ (possibly $(y^1,y^2)$ are defined only locally, therefore
we might need more charts).  The integral takes the form
\begin{equation*} \int d\phi \int d^{2}y f(y,\phi)\sqrt{E^3_i(y,\phi)E^3_i(y,\phi)} .\end{equation*}
The originally independent of $\phi$ smearing function $f$ becomes a function of $\phi$
via the choice of the coordinates.  Our point is, that 
\begin{equation*} d^{2}y \sqrt{E^3_i(y,{\phi})E^3_i(y,{\phi})}  \end{equation*}
is the area element of the surface $x^3=\phi={\rm const}$ with respect to the 3-geometry in $\Sigma$ 
corresponding to the densitized frame $E^a_i$.  In LQG that quantum area element is a well defined  operator distribution 
in ${\cal H}_\GR$. 
The only subtlety is,  that in our case the function $\phi$ is also a dynamical field, therefore we are talking about an operator
 \begin{equation*} \text{d}^3x \sqrt{\widehat{E}^a_i\widehat{E}^b_i\widehat{\phi}_{,a}\widehat{\phi}_{,b}}.  \end{equation*}         
In the very quantum representation of $\phi$, this is not a problem.
Given a quantum state 
\begin{equation*}\ket{\varphi}\otimes \Psi\in{\cal H}\ ,\end{equation*} 
$\varphi$ is a fixed function on $\Sigma$,  
therefore, the action of the quantum operator reads
\begin{equation}
\begin{split}\label{int} \int \text{d}^3 x &f(x) \sqrt{\widehat{\phi}_{,a}(x)\widehat{\phi}_{,b}(x)\otimes \widehat{E}^a_{i}(x)\widehat{E}^b_{i}(x)}\ \ket{\varphi}\otimes \Psi\\
 &= 
\ket{\varphi}\otimes  \int d^{3}x f(x)\sqrt{\widehat{E}^a_i(x)\varphi_{,a}(x)\widehat{E}^b_i(x)\varphi_{,b}(x)}\Psi. 
\end{split}
\end{equation}
The operator on the right hand side is well defined
on  the spin-network  cylindrical functions (\ref{sn}).  Indeed, this is a minor modification  of the Ma-Ling operator \cite{gr-qc/0005117} 
\begin{equation*}
\int \widehat{Q}(\omega)\ =\ \text{d}^3x \sqrt{\widehat{E}^a_i\widehat{E}^b_i\omega_a\omega_b}\end{equation*}
defined for an arbitrary differential 1-form $\omega$ on $\Sigma$. The  action of the Ma-Ling operator on a spin network state
(\ref{sn}) defined by a graph $\gamma=\{e_1,...,e_n\}$ of the edges colored by the spins $j_1,...,j_n$
is
\begin{equation*} \begin{split} \int \text{d}^3x &\sqrt{\widehat{E}^a_i\widehat{E}^b_i\omega_a\omega_b} \ket{\gamma,j,\iota}\\
&=\ 8\pi \beta \ell_\text{P}^2\sum_I
\sqrt{j_I(j_I+1)}\int_{0}^{1}dt |\omega_a\dot{e}_I^a(t)|\,\ket{\gamma,j,\iota}
\end{split}
\end{equation*}
where $e_I: [0,1]\rightarrow  \Sigma$ is a parametrization of the edge $e_I$, $I=1,...,n$. 
In our case $\omega=d\varphi$ what makes the derivation of the action of the operator yet simpler, therefore
we present it now. 
We calculate the operator first for  quantum states  such that  the the  graph $\{e_1,...e_n\}$ is entirely contained in 
a region of $\Sigma$ such that 
\begin{equation*}    d\varphi\ \not=\ 0\ .\end{equation*}
Then,
\begin{equation*} \begin{split}
\int d^{3}x &f(x)\sqrt{\widehat{E}^a_i(x)\varphi_{,a}(x)\widehat{E}^b_i(x)\varphi_{,b}(x)}\ket{\gamma,j,\iota}\\
& = \int d \varphi \int d^2 y f(y,\varphi)\sqrt{\widehat{E}^3_i(y,\varphi)(x)\widehat{E}^3_i(y,\varphi)}\ket{\gamma,j,\iota}. 
\end{split}
\end{equation*}
Every spin-network state $\ket{\gamma,j,i}$  is an eigenfunction of the generic quantum area element
$d^2 y f(y,\varphi)\sqrt{\widehat{E}^3_i(y,\varphi)\widehat{E}^3_i(y,\varphi)}$, hence  we obtain
\begin{equation*}\begin{split}  
\int  d\varphi \int d^2 y f(y,\varphi)&\sqrt{\widehat{E}^3_i(y,\varphi)\widehat{E}^3_i(y,\varphi)}\ket{\gamma,j,\iota}\\
&= (\int d\varphi a(\varphi)) \ket{\gamma,j,\iota} 
\end{split}
\end{equation*}
where the eigenvalue is computed below. 
For a generic value of $\varphi$ the constancy surface of $\varphi$ intersects 
each edge $e_I$ of the graph in finitely many (including $0$) isolated points, say
$(y_{I1}(\varphi),\varphi),...,(y_{Im_I}(\varphi),\varphi)$  and each
intersection point  contributes the value  $\ell_\text{P}f(y_{Ik},\varphi)\sqrt{j_{I}(j_{I}+1)}$, that is
\begin{equation}
\label{a} 
\begin{split}
a(\varphi) =   8\pi \beta \ell_\text{P}^2 \sum_I &\sqrt{j_{I}(j_{I}+1)}\cdot\\
&\cdot\left( f(y_{I1}(\varphi),\varphi)+\ldots+ f(y_{Im_I}(\varphi),\varphi) \right)  
\end{split}
\end{equation}
That formula holds as we change the value of $\varphi$, until the constancy surface
passes through either a vertex $v$ of $\gamma$ or the value at which the number of intersection
points with of the edges changes. Then, for that value of $\varphi$ the eigenfunction
takes a  different, still finite value. After passing that point, $a(\varphi)$ again takes the form
 (\ref{a}) with different intersection points.  The contribution to $\int d\varphi a(\varphi)$
 from each  edge $e_I$ is 
 \begin{equation*}\int_{e_I} f |d\varphi |\ =\ \int_{0}^{1}dt  f(e_I(t)) |\frac{d}{dt}\varphi(e_I(t))|\,dt\  .
 \end{equation*}
 In fact, the result
 is parametrization independent.   Subdivide $e_I$   into segments $e^{(1)}_I, ..., e^{(m_I)}_I$ such that
 $\varphi$ restricted to each of the segments is monotonic, and orient each of the segments
 such that the restriction of $\varphi$ is not decreasing along the segment.
 Then,  we obtain
 \begin{equation*}\int_{0}^{1}  f(e_I(t)) |\frac{d}{dt}\varphi(e_I(t))|\,dt\  =\ \sum_{k=1}^{m_I} \int_{e^{(k)}_I} f d\varphi .  
 \end{equation*}
 Finally,  
 \begin{equation} \label{theresult}
 \begin{split}
 \int d\varphi a(\varphi) &= 8\pi \beta \ell_\text{P}^2\sum_I\sqrt{j_I(j_I+1)}\int_{e_I} f|d\varphi|\\
 &=8\pi \beta \ell_\text{P}^2\sum_I\sqrt{j_I(j_I+1)}\sum_{k=1}^{m_I} \int_{e^{(k)}_I} fd\varphi  .
 \end{split}
 \end{equation}

The non-generic points do not contribute to the integral. Notice, that the case when
a segment of an edge overlaps a constancy surface of $\varphi$ is also contained
in the formula above, and contributes $0$ to the eigenvalue.  

For a general  state  $\ket{\varphi}\otimes\ket{\gamma,j,\iota}$, the manifold $\Sigma$ splits   into two disjoint regions  
\begin{equation*}\Sigma \ =\ \Sigma_1\cup\Sigma_2,\end{equation*}
where  $\Sigma_1$ is the maximal subset such that   
\begin{equation*} d\varphi |_{\Sigma_1}\not= 0.\end{equation*}  
Now,  $\Sigma_2$ does not contribute to  the integral (\ref{int})  
and the contribution from $\Sigma_1$  is given by   (\ref{theresult})
($\Sigma_1$ is in general disconnected, but on each connected component
we  repeat the calculation presented above).   

In conclusion, with the abbreviation $\ket{\varphi\gamma j \iota}:= \ket{\varphi}\otimes \ket{{\gamma,j,\iota}}$
\begin{widetext}
\begin{align}\label{theoperator}
\int \text{d}^3 x & f(x) \sqrt{\widehat{\phi}_{,a}(x)\widehat{\phi}_{,b}(x)\otimes \widehat{E}^a_{i}(x)\widehat{E}^b_{i}(x)}\ \ket{\varphi\gamma j \iota} =
 8\pi \beta \ell_\text{P}^2\left(\sum_I\sqrt{j_I(j_I+1)} \int_{e_I}f|d\varphi |\right) \ket{\varphi\gamma j \iota} \nonumber\\ 
 &=8\pi \beta \ell_\text{P}^2\left(\sum_I\sqrt{j_I(j_I+1)} \int_0^1dt f(e_I(t))|\frac{d\varphi(e_I(t))}{dt})|\right) \ket{\varphi\gamma j \iota}=
 8\pi \beta \ell_\text{P}^2\left(\sum_I\sqrt{j_I(j_I+1)}\sum_{k=1}^{m_I} \int_{e^{(k)}_I,} fd\varphi\right) \ket{\varphi\gamma j \iota},
\end{align}
\end{widetext}
for arbitrary spin-network function, where $I$ in the first sum on the right hand side ranges the labels
of the edges of $\gamma$, and  $e_I^{(1)},...,e_I^{(k)},..., e_I^{(m_I)}$ are segments of  $e_I$ such that 
$\varphi |_{e_I^{(k)}}$ is monotonic, and they are oriented in such a way that $\varphi$ is growing along each 
of them (notice that the integral of $\int _{e_I^{(k)}}d\varphi\cdot$ is sensitive on the orientation).                

A natural extension by the duality  of the operator is
\begin{widetext}
\begin{align}\label{theoperatordual}
&\left(\int \text{d}^3 x  f(x) \sqrt{\widehat{\phi}_{,a}(x)\widehat{\phi}_{,b}(x)\otimes \widehat{E}^a_{i}(x)\widehat{E}^b_{i}(x)}\right)^* 
\sum_\varphi \Psi[\varphi]\bra{\varphi}\otimes \bra{{\gamma,j,\iota}}
 = \sum_\varphi  \Psi'[\varphi] \bra{\varphi}\otimes \bra{{\gamma,j,\iota}}\nonumber\\
& \qquad\Psi'[\varphi]\ =\ 8\pi \beta \ell_\text{P}^2\left(\sum_I\sqrt{j_I(j_I+1)}\sum_{k=1}^{m_I} \int_{e^{(k)}_I,} fd\varphi\right) \Psi[\varphi] .
\end{align}
In particular, this extension applies also to the momentum representation and the polymer  representation!
\begin{equation}\label{theoperatordual'}
\left(\int \text{d}^3 x f(x) \sqrt{\widehat{\phi}_{,a}(x)\widehat{\phi}_{,b}(x)\otimes \widehat{E}^a_{i}(x)\widehat{E}^b_{i}(x)}\right)^* \bra{p}\otimes\bra{\gamma,j,\iota} =
 \sum_\varphi8\pi \beta \ell_\text{P}^2\left(\sum_I\sqrt{j_I(j_I+1)}\sum_{k=1}^{m_I} \int_{e^{(k)}_I,} fd\varphi e^{i\int p\varphi}\right)  \bra{\varphi}\otimes \bra{{\gamma,j,\iota}}
\end{equation}
\end{widetext}

\subsection{Extension to the case with nonzero potential}\label{A+B}
We turn now to the quantization of the part  of the constraint (\ref{C}) which includes also the potential term, that is 
of the classical expression ($f$ is an arbitrary smearing function)
$$ \int \text{d}^3x N\sqrt{\phi_{,a} E^a_i\phi_{,b}E^b_i  + 
2 u(\phi)|{\rm det}E|}(x) . $$

The operator $\widehat{\sqrt{{\rm det}{E}}}(x)$ defines the quantum 3-volume element in $\Sigma$
\begin{equation*} \widehat{V}(x)\text{d}^3 x\ =\ \widehat{\sqrt{{\rm det}{E}}}(x)\text{d}^3 x . \end{equation*}
Suitable choice of the tensor $\iota$ in (\ref{sn})  makes this cylindrical function an
eigenfunction of the volume element:
 \begin{equation*} \int \text{d}^3x N(x) \widehat{V}(x) \ket{\gamma,j,\iota}\ =\ \sum_{\alpha}f(v_\alpha)V_{v_\alpha}\ket{\gamma,j,\iota}  \end{equation*} 
for every function $N$, where $v_\alpha$, $\alpha=1,...,m$, are the vertices of $\gamma$,  $V_{v_\alpha}$ are suitable eigenvalues depending on the colorings $j$ by spins  and $\iota$ by the tensors. There is a basis of such cylindrical functions. 
The coupling with $u(\widehat{\phi(x)})$ is also automatic 
\begin{equation*} 
\begin{split}
\int \text{d}^3x N(x)& \sqrt{u(\widehat{\phi})}\widehat{V}(x)\ket{\varphi}\otimes \ket{\gamma,j,\iota}\\ 
&= \sum_{\alpha}N(v_\alpha)\sqrt{u(\varphi(v_\alpha))}V_{v_\alpha} \ket{\varphi}\otimes \ket{\gamma,j,\iota}  
\end{split}
\end{equation*} 

So, we already have in our framework the well defined  two quantum operators
\begin{equation*}
\int \text{d}^3xN(x)\sqrt{\widehat{A}(x)}, \ \ \ \ \ \ \ {\rm and} \ \ \ \ \ \ \ \int \text{d}^3xN(x)\sqrt{\widehat{B}(x)},
\end{equation*}
corresponding to the following classical expressions:
\begin{align} \label{AB}
A(x) \ &:=\  {\phi}_{,a}(x) {E}^a_i(x){\phi}_{,b}(x){E}^b_i(x)\\
B(x)   \ &:=\  2 u(\phi(x))|{\rm det}E(x)| .
\end{align}
What we need is to define the operator
\begin{equation*}
\int \text{d}^3 x f(x) \sqrt{\widehat{A}(x)+\widehat{B}(x)}.
\end{equation*}
To this end, we introduce a parametrized by $\epsilon$ family of  partitions of $\Sigma$
$$ \Sigma\ =\ \bigcup_r\Sigma^\epsilon_r  ,$$  
such that the cells  $\Sigma^\epsilon_r$ are shrank uniformly as $\epsilon\rightarrow 0$.
Now, the trick is to notice that
\begin{widetext}
\begin{equation*} 
\int_\Sigma \text{d}^3 x f(x) \sqrt{A(x) + B(x)}\ =\ {\rm lim}_{\epsilon\rightarrow 0} 
\sum_r  \sqrt{\left(\int_{\Sigma^\epsilon_r} \text{d}^3 x f(x)\sqrt{A(x)}\right)^2 + \left(\int_{\Sigma\epsilon_r} \text{d}^3 x f(x)\sqrt{B(x)}\right)^2}
\end{equation*}
The proposal is to define 
\begin{align*} 
\int_\Sigma \text{d}^3 x f(x) \sqrt{\widehat{A}(x) + \widehat{B}(x)}\ \ket{\varphi\gamma j \iota} :=
 {\rm lim}_{\epsilon\rightarrow 0} 
\sum_r  \sqrt{\left(\int_{\Sigma^\epsilon_r} \text{d}^3 x f(x)\sqrt{\widehat{A}(x)}\,\right)^2 +
\left(\int_{\Sigma\epsilon_r} \text{d}^3 x f(x)\sqrt{\widehat{B}(x)}\,\right)^2}\ \ket{\varphi\gamma j \iota}
\end{align*}
Let us discuss the properties of the  individual operators  
$$\int _{\Sigma^\epsilon_r}f(x)\sqrt{\widehat{A}(x)}\text{d}^3 x,\ \ \ \  {\rm and}\ \ \ \  \int _{\Sigma^\epsilon_r}\text{d}^3 x f(x)\sqrt{\widehat{B}(x)}\text{d}^3 x,$$
respectively,  when the domain $\Sigma^\epsilon_r$ of integration is shrank to a point.  
For both of them each state of the form $\ket{\varphi}\otimes\ket{\gamma,j,\iota}$ is an eigenstate. Denote the eigenvalues
by $\lambda_A^\epsilon{}_r$, and, respectively $\lambda_B^\epsilon{}_r$. It follows, that
\begin{equation*} 
\sum_r  \sqrt{\left(\int_{\Sigma^\epsilon_r} \text{d}^3 x f(x)\sqrt{\widehat{A}(x)}\,\right)^2 +
\left(\int_{\Sigma\epsilon_r} \text{d}^3 x f(x)\sqrt{\widehat{B}(x)}\,\right)^2}\ \ket{\varphi}\otimes\ket{\gamma,j,\iota}\ =\
\sum_r \sqrt{(\lambda_A^\epsilon{}_r)^2 + (\lambda_B^\epsilon{}_r)^2} \ \ket{\varphi}\otimes\ket{\gamma,j,\iota}.   
\end{equation*}
\end{widetext}

In the first case, we have ($e_1,...,e_n$ and $v_1,...,v_m$, are the edges and, respectively, vertices of $\gamma$)
\begin{align*}
\lambda_A^\epsilon{}_r\ &=\ 8\pi \beta \ell_\text{P}^2\sum_{I}\int_{e^\epsilon_I} | f d\varphi |\sqrt{j_I(j_I+1)}\\
e^\epsilon_I\ &:=\ e_I\cap \Sigma^\epsilon_r
\end{align*} 
Respectively, in the second case
\begin{align*}  \lambda_B^\epsilon{}_r\ & =\ 
\sum_{\alpha} f(v_\alpha)\sqrt{u(\varphi(v_\alpha))}V_{v_\alpha}\\
v_\alpha\ &\in \Sigma^\epsilon_r\ .
\end{align*} 
For sufficiently fine partition, each cell $\Sigma^\epsilon_r$ either does not contain any vertex $v_\alpha$, and then
$$ \lambda_B^\epsilon{}_r\ =\ 0$$
or, $\Sigma^\epsilon_{r_\alpha}$ contains exactly one vertex $v_{\alpha}$, and then
$$ \lambda_B^\epsilon{}_{r_\alpha}\ =\ f(v_\alpha)\sqrt{u(\varphi(v_\alpha))}V_{v_\alpha}$$
independently of $\epsilon$.
On the other hand,   
$$ \lambda_A^\epsilon{}_{r_\alpha}\ \rightarrow\ 0,$$
as $\epsilon\ \rightarrow\ 0$. 

In conclusion 
\begin{align*} 
{\rm lim}_{\epsilon\rightarrow 0} &\sum_r \sqrt{(\lambda_A^\epsilon{}_r)^2 + (\lambda_B^\epsilon{}_r)^2} \\   
&=\sum_r \lambda_A^\epsilon{}_r  \ +\ \sum_{\alpha}  \lambda_B^\epsilon{}_{r_\alpha} ,
\end{align*}
where the right hand side is already $\epsilon$ invariant. Finally, due to the different characters of the operators $\widehat{A}(x)$
and, respectively, $\widehat{B}(x)$ with respect to the partitioning $\Sigma$, in this case of (\ref{AB}) and their quantizations, the result reads
\begin{widetext}
\begin{equation*}
\int_\Sigma \text{d}^3 x f(x)\sqrt{\widehat{A}(x)+\widehat{B}(x)}\ =\   \int_\Sigma \text{d}^3 x f(x)\sqrt{\widehat{A}(x)} +\int_\Sigma \text{d}^3 x f(x)\sqrt{\widehat{B}(x)}  .
\end{equation*}
That is, our final result is
\begin{align*} 
\int \text{d}^3x N(x)&\sqrt{ \widehat{\phi}_{,a}(x) \widehat{E}^a_i(x)\widehat{\phi}_{,b}(x)\widehat{E}^b_i(x)  + 2 
u(\widehat{\phi}(x))|{\rm det}\widehat{E}(x)|}\,\ket{\varphi}\otimes\ket{\gamma,j,\iota}\ = \nonumber\\ 
&\qquad\qquad\qquad\qquad \left(8\pi \beta \ell_\text{P}^2\sum_{I}\int_{e_I}N|d\varphi|\sqrt{j_I(j_I+1)}\ +  \sum_{\alpha} N(v_\alpha)\sqrt{u(\varphi(v_\alpha))}V_{v_\alpha}\right)\, \ket{\varphi}\otimes\ket{\gamma,j,\iota}.
\end{align*}

A peculiar property of our result is the following quantum identity     
\begin{align*} 
\int \text{d}^3x f(x)&\sqrt{(\widehat{\phi}_{,a}(x) \widehat{E}^a_i(x)\widehat{\phi}_{,b}(x)\widehat{E}^b_i(x)  + 2
u(\widehat{\phi}(x))|{\rm det}\widehat{E}(x)|} \nonumber\\ 
&\quad=\int \text{d}^3x f(x)\sqrt{\widehat{\phi}_{,a}(x) \widehat{E}^a_i(x)\widehat{\phi}_{,b}(x)\widehat{E}^b_i(x)}  + 
\int \text{d}^3x f(x)\sqrt{2  u(\widehat{\phi}(x))|{\rm det}\widehat{E}(x)|}
\end{align*}
which does not hold in the classical theory. 
\end{widetext}
\subsection{The operator  $\int \widehat{\pi}(x)f(x)$}
\label{pi_dual}
Actually, on the dual states we can also define an operator $\int \widehat{\pi}(x)f(x)$ 
itself
\begin{equation*}
\begin{split}
&\left(\left(\int \widehat{\pi}(x)f(x)\right)^*\gbra{\psi}\right)\ \ket{\varphi'}\\
&:=\ i\frac{d}{d\epsilon} 
\left(\exp\left(-i\int d^x \widehat{\pi}(x)f(x)\right)^*\sum_{\varphi}\psi[\varphi]\scpr{\varphi}{\varphi'}\right)\\ 
&=\  i\frac{d}{d\epsilon}  \psi[\varphi'+\epsilon f] =:i \delta_f \psi[\varphi']
\end{split}
\end{equation*}
where we have introduced the directional functional derivative $\delta_f$, and used notation defined in 
\eqref{generalized}. Hence
\begin{equation*}
\left(\int \widehat{\pi}(x)f(x)\right)^* \gbra{\psi}\ =\ i\,\gbra{\delta_f\, \psi}
\end{equation*}

\subsection{The quantization of  $\widehat{C}_\GR$ in ${\cal H}_\SF\otimes{\cal H}_\GR$}
\label{Cgr}
For an operator defined directly in the kinematical Hilbert space ${\cal H}_\GR$, say for 
$\widehat{V}[N]\ =\ \int_\Sigma \text{d}^3x N(x)\widehat{V}(x),$
the extension to the  Hilbert space ${\cal H}_{\rm matter}\otimes{\cal H}_\GR$ is straightforward, just
$${\rm id}\otimes\widehat{V}[N].$$ 
One even skips the tensor product and simply writes 
``$\widehat{V}[N]$''.
The gravitational part of the scalar constraint operator $\widehat{C}_\GR(N)$, on the other hand,
is not defined directly in the kinematical Hilbert space ${\cal H}_\GR$.  Instead, it is defined 
on partial solutions to the (matter free) diffeomorphism constraint, namely on states invariant with
respect to the diffeomorphisms  Diff$_{{\rm Vert}(\gamma)}$ preserving the vertices of a given graph 
$\gamma$. The partial solutions  are obtained by a suitable rigging map defined more precisely in
\cite{arXiv:1410.5276}
\begin{equation*}
\eta_\GR: \cyl_\GR \rightarrow\ \cyl_\GR^* . 
\end{equation*}
The rigging map is defined using some orthogonal decomposition
\begin{equation}\label{thedecomp}\cyl_\GR\ =\ \bigoplus_{\gamma}{\cal H}_\gamma
\end{equation}
where each $\gamma$  runs through the set  of  un-oriented graphs in $\Sigma$  and ${\cal H}_{\gamma}$ is constructed from suitably selected cylindrical functions
depending on the parallel transports  along the edges $e$ of $\gamma$.     
For every $\gamma$ and a spin-network $\ket{\gamma,j,\iota}$,
the result $\eta_\GR(\ket{\gamma,j,\iota})$ is Diff$_{{\rm Ver}(\gamma)}$ invariant. 

In ${\cal H}_{\rm matter}\otimes{\cal H}_\GR$, however, in general  the diffeomorphism  invariance  
couples the two Hilbert spaces non-trivially.    
This problem does not occur  in the case when  ${\cal H}_{\rm matt}={\cal H}_{\rm Pol}$.
Indeed, given a state 
$$ \ket{p}\otimes\ket{\gamma,j,\iota}\ \in\ {\cal H}_{\rm Pol}\otimes {\cal H}_\GR,$$
such that $p=p(v_1)\delta(x,v_1)+...+p(v_m)\delta(x,v_m)$ and $v_1,...,v_m \in {\rm Ver}(\gamma)$,
the state $\ket{p}$ is invariant with respect to the Diff$_{v_1,...,v_m}$. Hence,  
the state
$$ \ket{p}\otimes \eta_\GR(\ket{\gamma,j,\iota})$$
is a partial solution to the {\it total} diffeomorphism contraint
$$ \left(U_f\otimes U_f\right)\, \left(\ket{p}\otimes \eta_\GR(\ket{\gamma,j,\iota})\right)\ =\ \ket{p}\otimes \eta_\GR(\ket{\gamma,j,\iota})
$$
for every $f\in$Diff$_{v_1,...,v_m}$. On such states, the LQG part of the scalar constraint is defined just
as
$$ {\rm id}\otimes \widehat{C}_\GR(N).$$
A similar observation applies if given $\ket{\gamma,j,\iota}\in {\cal H}_\gamma$ as above, instead of $\ket{p}$  we take     
a generalized state of the position representation which has the form
$$ \sum_{\varphi}\Psi(\varphi(v_1),...,\varphi(v_m))\bra{\varphi}.$$
The latter state is Diff$_{{\rm Ver}(\gamma)}$ invariant, therefore again
\begin{equation*}
\begin{split}
&\left(U_f\otimes U_f\right)  \sum_{\varphi}\Psi(\varphi(v_1),\ldots,\varphi(v_m))\bra{\varphi} \otimes \eta_\GR(\ket{\gamma,j,\iota})\\
&= \sum_{\varphi}\Psi(\varphi(v_1),\ldots,\varphi(v_m))\bra{\varphi}\otimes \eta_\GR(\ket{\gamma,j,\iota})
\end{split}
\end{equation*}
for every $f\in$Diff$_{v_1,...,v_m}$ and the simple product operator
$$ {\rm id}\otimes \widehat{C}_\GR(N)$$
is the natural  definition of the matter free part of the scalar constraint
on a state
\begin{equation*}
 \sum_{\varphi}\Psi(\varphi(v_1),...,\varphi(v_m))\bra{\varphi}\otimes \eta_\GR(\ket{\gamma,j,\iota}).
\end{equation*}  

However, for a state 
$$ \ket{\varphi}\otimes\ket{\gamma,j,\iota}\ \in\ {\cal H}_\SF\otimes{\cal H}_\GR$$
that straightforward extension of $\widehat{C}_\GR(N)$ to ${\rm id}\otimes \widehat{C}_\GR(N)$
does not apply. Now we describe a solution of that issue valid for a class of states $\ket{\varphi}$. 

For every graph $\gamma$ in the decomposition (\ref{thedecomp}) consider  states $\ket{\varphi}$ that satisfy the following assumption: 
\begin{itemize}
\item[A:] {\it There is a neighborhood of ${\rm Vert}(\gamma)$ in which the function $\varphi$  is constant.} 
\end{itemize}

Given a space ${\cal H}_{\gamma}$ of the decomposition   (\ref{thedecomp}), and a state $\ket{\varphi}$
which satisfies the assumption A, we consider a subgroup Diff$_{\varphi,{\rm Ver}(\gamma)}$ of
Diff defined as follows
\begin{equation}
\label{thediffs}{\rm Diff}_{\varphi,{\rm Ver}(\gamma)}\ :=\ \{ f\in {\rm Diff}\ :\ f_{{\rm Ver}(\gamma)} = {\rm id}, \ \ \ \ f^*\varphi=\varphi\}.  
\end{equation}    
That is, those diffeomorphisms preserve the vertices of $\gamma$  and otherwise act freely in the subsets of constancy of 
$\varphi$.  
Next, with the group ${\rm Diff}_{\varphi,{\rm Ver}(\gamma)}$,  we average   elements of $\ket{\varphi}\otimes{\cal H}_\gamma$ 
To this end, we introduce the subgroup  
\begin{equation*} {\rm TDiff}_{\varphi,\gamma}\ :=\ \{f\in {\rm Diff}\ :\ U_f|_{\ket{\varphi}\otimes{\cal H_\gamma}}\ =\ {\rm id}\} 
\end{equation*} 
consisting of all the diffeomorphisms of $\Sigma $ which act trivially in the given $\ket{\varphi}\otimes{\cal H}_\gamma$. 
The averaging map $\eta$ is defined as follows
\begin{equation*}
\ket{\varphi}\otimes{\cal H}_\gamma\ni \ket{\varphi}\otimes\ket{\gamma,j,\iota}\ \mapsto\  \bra{\varphi}\otimes \eta(\ket{\gamma,j,\iota}) 
\end{equation*}
with 
\begin{equation*}
 \eta(\ket{\gamma,j,\iota})=\frac{1}{n_{\varphi,\gamma}}\sum_{[f]}\bra{\varphi}\otimes(U_f)^*\bra{\gamma,j,\iota}\ \in\ {\cyl}_\GR^*
 \end{equation*}
and $[f]\in {\rm Diff}_{\varphi,{\rm Ver}(\gamma)}\,/\,
{\rm TDiff}_{\varphi,\gamma} $. $n_{\varphi,\gamma}$ is an arbitrarily fixed number. 
The elements of the resulting space $\bra{\varphi}\otimes\eta({\cal H}_\gamma)$ are  Diff$_{\varphi,{\rm Ver}(\gamma)}$ invariant, hence 
they are solutions of the  quantum vector   constraints corresponding to all the shift vectors  
$$\vec{N}(v_1)=\ \ldots\ \vec{N}(v_m)\ =\ \vec{N}_{\Sigma\setminus{\rm Supp}(\varphi)}\ =\ 0  .$$
   
By linearity we combine all the $\eta$ to a single averaging map
\begin{equation*}
 \eta :  \cyl_\SF\otimes\cyl_\GR\ \rightarrow\ \left(\cyl_\SF\cyl_\GR\right)^* . 
 \end{equation*}
In the space   $\eta\left(\cyl_\SF\otimes\cyl_\GR\right)$ the Hilbert product is defined 
by
\begin{equation*}
(\eta(\Psi)|\eta(\Psi'))\ :=\ \eta(\Psi)(\Psi') ,
\end{equation*}
and makes it a Hilbert space upon the completion.    
For every finite set $\{v_1,...,v_m\}\subset \Sigma$, and for every $\ket{\varphi}$ such that  
$$ d\varphi|_{\cal U} = 0,$$  
for some neighborhood ${\cal U}$ of $\{v_1,...,v_m\}$, consider all the graphs $\gamma$ used in the decomposition 
(\ref{thedecomp}) such that
\begin{equation}\label{gammavvv} {\rm Ver}(\gamma)\ =\ \{v_1,...,v_m\}, \end{equation}
and the corresponding space
\begin{equation*}
{\cal H}_{\varphi,\{v_1,...,v_m\}}\ :=\ \eta\left(\ket{\varphi}\otimes\bigoplus_{\gamma} {\cal H}_\gamma \right)
\end{equation*}
where the summation is with respect to the  $\gamma$s such that (\ref{gammavvv}) holds.
We have the direct decomposition
\begin{equation}\label{Hnew}
\eta\left(\cyl_\SF\otimes\cyl_\GR\right)\ =\ \bigoplus_{\varphi,\{v_1,...,v_m\}\subset\Sigma}
{\cal H}_{\varphi,\{v_1,...,v_m\}} 
\end{equation} 
  
The derivation of the quantum scalar constraint operator presented in \cite{arXiv:1410.5276} is valid in 
$\eta\left(\cyl_\SF\otimes\cyl_\GR\right)$ after a minor restriction  on the regulated operator 
$$ \widehat{C}^\epsilon_\GR(N):\cyl_\GR \rightarrow  \cyl_\GR\, :$$
namely, given a state $\ket{\varphi}\otimes\ket{\gamma,j,\iota}$ the loops added by the operator
at the vertices of $\gamma$ should be all contained in the constancy neighborhoods of a given $\varphi$. 
Since the loops are shrank to the vertices anyway, the restriction does change the resulting limit.           
Then, given a state $ \eta\left(\ket{\varphi}\otimes\ket{\gamma,j,\iota}\right) $ the dual action of
the operator  
$$  \left({\rm id}\otimes \widehat{C}^\epsilon_\GR(N)\right)^* \left(\bra{\varphi}\otimes \bra{\gamma,j,\iota}\right) $$
is independent of $\epsilon$ and defines an operator 
\begin{equation}\label{idotimesC}
{\rm id}\otimes \widehat{C}_\GR(N)\ :\ \eta\left(\cyl_\SF\otimes\cyl_\GR\right)
\rightarrow\ \eta\left(\cyl_\SF\otimes\cyl_\GR\right) .
\end{equation}
The operator preserves the orthogonal decomposition (\ref{Hnew})
\begin{equation}\label{idotimesC'}
{\rm id}\otimes \widehat{C}_\GR(N)\ :\ {\cal H}_{\varphi,\{v_1,...,v_m\}} \ 
\rightarrow\ {\cal H}_{\varphi,\{v_1,...,v_m\}}  .
\end{equation} 
\section{Solving toy models of the quantum constraint}
In this section, as a warm-up exercise to one day solve the full constraint, we find solutions 
to the equation
\begin{equation*} 
 \left( \widehat{\pi}(x) + \widehat{h}(x) \right)\Psi\ =\ 0  
 \end{equation*}
with various examples of operator $\widehat{h}(x)$. The solutions give clues 
about the class of states on which the gravitational constraint can be quantized. 

The simplest case is $h=0$. Here the solution 
is, according to section \ref{pi_dual}, simply the dual state $\gbra{\Psi}$ with $\Psi[\varphi]=c=\text{const.}$
\begin{equation*}
\pi(h)^*\gbra{c}\equiv\gbra{i\delta_h c}=0
\end{equation*}
We will now turn to some more complicated cases.
\subsection{States annihilated by $\widehat{\pi}(x) + u(\widehat{\phi}(x))\widehat{\sqrt{{\rm det}{E}}}(x)$}
Let us now consider  in ${\cal H}_\SF\otimes{\cal H}_\GR$ 
an equation
\begin{equation} \label{pi+uV}
\left(\widehat{\pi}(x) + u(\widehat{\phi}(x))\widehat{\sqrt{{\rm det}{E}}}(x)\right)^*  \gbra{\Psi} \otimes  \bra{\gamma, j, \iota} \ =\ 0. 
\end{equation}
This equation 
amounts to 
\begin{equation}
\label{pi+uV'} i\frac{d}{d\epsilon}\Psi[\varphi + \epsilon f] + \sum_{x\in \text{Ver}(\gamma)} f(x) V_{x}u(\varphi(x))\Psi[\varphi] \ =\ 0. 
\end{equation} 
Substitute for $\Psi$ in (\ref{pi+uV'}) : 
\begin{equation*} 
\Psi[\varphi]\ =\ \exp \left(i\ \sum_{x\in \text{Ver}(\gamma)}V_{x}\int_{\varphi_0(x)}^{\varphi(x)}\,\text{d} \varphi'(x) u(\varphi'(x))\right)\Psi_0[\varphi]
\end{equation*} 
where $\varphi_0:\Sigma\rightarrow \mathbb{R}$ is  an arbitrarily fixed function (constant of integration). 
Then (\ref{pi+uV'}) is equivalent to
\begin{equation*}
\frac{d}{d \epsilon}\Psi_0[\varphi+\epsilon f]\ =\ 0,
\end{equation*}
hence 
\begin{equation*}\psi_0(\varphi)\ =\ {\rm const} .\end{equation*}      
In conclusion, the general solution to the equation (\ref{pi+uV}) is $ \gbra{\Psi_{\varphi_0}} \otimes  \bra{\gamma, j, \iota}$ with
\begin{equation*}
\Psi_{\varphi_0}[\varphi]=\exp \left(i\,\sum_{x\in \text{Ver}(\gamma)}V_{x}\int_{\varphi_0(x)}^{\varphi(x)}\,\text{d}\varphi'(x) 
u(\varphi'(x))\right).
\end{equation*}
It can be written in the compact form
\begin{equation*}
\exp\left(i\,\int \text{d}^3x\, \int_{\varphi_0(x)}^{\widehat{\phi}(x)}{\text{d}\varphi}'(x)\,
u({\varphi}'(x))\otimes \widehat{V}(x)\right)^*\gbra{\Psi} \otimes  \bra{\gamma, j, \iota}
\end{equation*}
The general solution to the equation (\ref{pi+uV}) is 
\begin{equation*}
\Psi\ =\ 
\sum_{\varphi_{0},\gamma,j,\iota}\alpha_{\varphi_0,\gamma,j,\iota}\gbra{\Psi_{\varphi_0}} \otimes  \bra{\gamma, j, \iota}.
\end{equation*}

\subsection{States annihilated by $\int \text{d}^3 x (\widehat\pi +a\sqrt{\widehat{E}^a_i\widehat{E}^b_i\widehat{\phi}_{,a}\widehat{\phi}_{,b}})$}
The next equation of the considered form for which we have an elegant solution is again imposed
on the states $ \gbra{\Psi} \otimes  \bra{\gamma, j, \iota}$:
\begin{equation}\label{piaEEphiphi}
\int \text{d}^3x \left(\widehat\pi(x)\ +\ a \sqrt{\widehat{E}^a_i\widehat{E}^b_i\widehat{\phi}_{,a}\widehat{\phi}_{,b}(x)}\right)  \gbra{\Psi}\otimes\bra{\gamma,j,\iota}\ =\ 0 .
\end{equation}
The action of the first operator  reads
\begin{equation}\label{first}
\int \text{d}^3x \widehat\pi(x)\gbra{\Psi}\otimes \bra{\gamma,j,\iota}\ = \  i\gbra{\delta_1\Psi}\otimes \bra{\gamma,j,\iota}
\end{equation}
whereas the action of the second one is
\begin{align}\label{second}
\int \text{d}^3x\sqrt{\widehat{E}^a_i\widehat{E}^b_i\widehat{\phi}_{,a} \widehat{\phi}_{,b}(x)}\gbra{\Psi}\otimes \bra{\gamma,j,\iota} \ =\ \gbra{\lambda'_{\gamma,j}\Psi}\otimes \bra{\gamma,j,\iota}\nonumber\\
\lambda'_{\gamma,j}[\varphi]\ :=\ 8\pi \beta \ell_\text{P}^2 \left(\sum_I\sqrt{j_I(j_I+1)}\sum_{k=1}^{m_I} \left(\varphi(e_{I+}^{(k)})-\varphi(e_{I-}^{k})\right)\right) 
\end{align} 
where the beginning/end of each segment $e_I^{(k)}$ is denoted by $e^k_{I-/+}$. As suggested by the notation, we have 
\begin{align*}
\lambda'_{\gamma,j}[\varphi]\ &= \delta_1\lambda'[\varphi]_{\gamma,j} = \frac{d}{d\epsilon}|_{\epsilon=0}\lambda(\varphi+\epsilon)_{\gamma,j}\nonumber\\ 
\lambda_{\gamma,j}[\varphi]\ &=\ 
\frac{1}{2}8\pi \beta \ell_\text{P}^2 \left(\sum_I\sqrt{j_I(j_I+1)}\sum_{k=1}^{m_I} \left(\varphi(e_{I+}^{(k)})^2-\varphi(e_{I-}^{k})^2\right)\right)
\end{align*}
The comparison between (\ref{first}) and (\ref{second})  leads us to the operator  
 \begin{equation*} 
 \ket{\varphi}\otimes\ket{\gamma,j,\iota}\ \mapsto\   \lambda(\varphi)_{\gamma,j}\ket{\varphi}\otimes\ket{\gamma,j,\iota} . 
 \end{equation*}
This operator can be written in the compact form 
\begin{equation*}
\int \text{d}^3x\,\widehat{\phi}(x)\sqrt{\widehat{E}^a_i\widehat{E}^b_i\widehat{\phi}_{,a}\widehat{\phi}_{,b}(x)} .
\end{equation*}
Indeed, the action of the latter one on a state $\ket{\varphi}\otimes \ket{\gamma,j,\iota} $ coincides with the action of the operator
(\ref{theoperator}) with the function $\varphi$ substituted for an arbitrary smearing function $f$,
\begin{align*}
\int \text{d}^3x&\,\widehat{\phi}\sqrt{\widehat{E}^a_i\widehat{E}^b_i\widehat{\phi}_{,a}\widehat{\phi}_{,b}}\,\ket{\varphi\gamma j \iota}\\
&=8\pi \beta \ell_\text{P}^2 \left(\sum_I\sqrt{j_I(j_I+1)}\sum_{k=1}^{m_I} \int_{e^{(k)}_I,} \varphi \text{d}\varphi\right)\ket{\varphi\gamma j \iota}\\
&=\frac{1}{2}8\pi \beta \ell_\text{P}^2 \left(\sum_I\sqrt{j_I(j_I+1)}\sum_{k=1}^{m_I} \left(\varphi(e_{I+}^{(k)})^2-\varphi(e_{I-}^{k})^2\right)\right)
\ket{\varphi\gamma j \iota}\\
&=\lambda(\varphi)_{\gamma,j}\,\ket{\varphi\gamma j \iota}.
\end{align*}
The operator we actually need is the exponential
\begin{align*}
\exp&\left(ia\int \text{d}^3x\,\widehat{\phi}\sqrt{\widehat{E}^a_i\widehat{E}^b_i\widehat{\phi}_{,a}\widehat{\phi}_{,b}}\right)\,\ket{\varphi}\otimes \ket{\gamma,j,\iota}\\
&\qquad= e^{ia\lambda(\varphi)_{\gamma,j}}\,\ket{\varphi}\otimes \ket{{\gamma,j,\iota}} .
\end{align*}
Extended by the duality to the states  $\gbra{\Psi} \otimes  \bra{\gamma, j, \iota}$, the action is
\begin{align}\label{subst1}
\int \text{d}^3x\widehat{\phi}\sqrt{\widehat{E}^a_i\widehat{E}^b_i\widehat{\phi}_{,a}\widehat{\phi}_{,b}} 
\gbra{\Psi,\gamma,j,\iota}\nonumber &=\gbra{\lambda_{\gamma,j}\Psi,\gamma,j,\iota},\nonumber\\
\exp\left(ia\int \text{d}^3x\widehat{\phi}\sqrt{\widehat{E}^a_i\widehat{E}^b_i\widehat{\phi}_{,a}\widehat{\phi}_{,b}} \right)
\gbra{\Psi,\gamma,j,\iota}&=\bra{e^{ia\lambda_{\gamma,j}}\psi,\,\gamma,\,j,\,\iota}
\end{align}
(the latter equality can be also obtained by direct exponentiation of the former one). The key identity, the operators satisfy is
\begin{widetext}
\begin{equation*}
e^{ia\int \text{d}^3x\,\widehat{\phi}\sqrt{\widehat{E}^a_i\widehat{E}^b_i\widehat{\phi}_{,a}\widehat{\phi}_{,b}}}
\left(\int \text{d}^3x\,\widehat{\pi} \right) 
e^{-ia\int \text{d}^3x\,\widehat{\phi}\sqrt{\widehat{E}^a_i\widehat{E}^b_i\widehat{\phi}_{,a}\widehat{\phi}_{,b}} }\gbra{\Psi,\gamma,j,\iota}\ =\  \int \text{d}^3x \,\left( \widehat{\pi} + a\sqrt{\widehat{E}^a_i\widehat{E}^b_i\widehat{\phi}_{,a}\widehat{\phi}_{,b}}\right)\gbra{\Psi,\gamma,j,\iota}.
\end{equation*}
\end{widetext}
It follows that a general solution to the equation \eqref{piaEEphiphi} is 
\begin{equation*}
\gbra{\Psi,\gamma,j,\iota}\ =\ e^ {i a \int \text{d}^3x\,\widehat{\phi}\sqrt{\widehat{E}^a_i\widehat{E}^b_i\widehat{\phi}_{,a}\widehat{\phi}_{,b}}}
\bra{\Psi_0,\gamma,j,\iota}
\end{equation*}
where $\varphi\mapsto \Psi_0[\varphi]$, is an arbitrary functional which satisfies the following condition:
\begin{equation}
\label{subst2}
\frac{d}{d \epsilon} \psi_0(\varphi +\epsilon)\ =\ 0\ . 
\end{equation}

\section{The complete constraint}
\subsection{Putting everything together}
The advantage of the Hilbert space (\ref{Hnew}) we have introduced in the previous section, is that it 
\begin{itemize} 
\item admits the 
action of each ingredient of the right hand side of  (\ref{C})
\item is preserved by that action. 
\end{itemize}
Indeed, to start with, we have the operator ${\widehat{C}(N)}_\GR$, which preserves the decomposition  (\ref{idotimesC}).

Secondly, every state  $\eta\left(\ket{\varphi}\otimes\ket{\gamma,j,\iota} \right)$ is an eigenstate of the operator
(\ref{theoperatordual}), namely
 \begin{align}\label{theoperatordual''}
&\left(\int \text{d}^3 x N(x) \sqrt{\widehat{\phi}_{,a}(x)\widehat{\phi}_{,b}(x)\otimes \widehat{E}^a_{i}(x)\widehat{E}^b_{i}(x)}\right)^* \eta\left(\ket{\varphi}\otimes \ket{{\gamma,j,\iota}}\right)\nonumber\\ =  &\ell^2_{P}\left(\sum_I \sqrt{j_I(j_I+1)}\int_{e_I}N|d\varphi|\right)\, \eta\left(\ket{\varphi}\otimes \ket{{\gamma,j,\iota}}\right).
\end{align}
A subtlety is, that given a graph $\gamma$, in the state $\eta\left(\ket{\varphi}\otimes \ket{{\gamma,j,\iota}}\right)$
a part of the graph is unchanged, however a part of the graph is averaged with respect to the diffeomorphisms
(\ref{thediffs}).   In this way, for every edge  $e_I$ of $\gamma$, its two  parts, say $e^+_I$ and  $e^-_I$,  are affected 
by $\eta$.  Nonetheless, for those parts we have 
$$ \varphi_{e^\pm_I}\ =\ {\rm const}, $$ 
hence,   
$$ \int_{e^\pm_I} N |d\phi| =0$$ 
 and they do not contribute to the action of the operator (in a general case of averaging with respect to the diffeomorphisms, the diffeomorphisms act simultaneously on $\gamma$ and $\varphi$, preserving the integrals
 $\int_{e_I} |d\varphi|$, however the smearing function $N$ breaks the invariance).  
  
Next, we need an operator 
$$\sqrt{u(\widehat{\phi}(x))\otimes {|{\rm det}\widehat{E}(x)|}\ +\  \sqrt{|{\rm det}\widehat{E}(x)|}\widehat{C}_\GR(x)}$$
that has not beed defined yet. After the restriction  to a subspace ${{\cal H}_{\varphi,\{v_1,...,v_m\}}}$ ,
the $\widehat{\phi}$ operator turns into the function $\varphi$.
   Our task boils down  to the following problem: 
   
\noindent there are two operator valued
distributions $\widehat{A}(x)$ and $\widehat{B}(x)$, what we need is an operator
$\sqrt{\widehat{A}(x)\widehat{B}(x)}$. In our case, 
\begin{align*}\widehat{A}(x)\ &=\ {\rm id}\otimes \sqrt{|{\rm det}\widehat{E}(x)|}, \\ 
\widehat{B}(x)\ &=\   {\rm id}\otimes u(\varphi(x))\sqrt{|{\rm det}\widehat{E}(x)|} + \widehat{C}_\GR(x).
\end{align*}
The operators  restricted to each subspace
${\cal H}_{\varphi,\{v_1,...,v_m\}}$ satisfy
\begin{align*}
\widehat{A}(x)\ =\ {\rm id}\otimes{} \sum_{\alpha=\alpha}^m\delta(x,v_\alpha)\widehat{A}_{v_\alpha}, \ \ \ \ \ \widehat{B}(x)\ =\ {\rm id}\otimes\sum_{\alpha=1}^m\delta(x,v_\alpha)
\widehat{B}_{v_\alpha} .
\end{align*}
Specifically 
$$\widehat{A}_{v_\alpha}\ =\ \widehat{V}_{v_\alpha},\ \ \ \ \ \ \ \widehat{B}_{v_\alpha}=u(\varphi(v_\alpha))\widehat{V}_{v_\alpha}+
\widehat{C}_{\GR v_\alpha}.$$

We solve the problem as follows.   
We smear each of the operators against a smooth distribution $\delta^\epsilon(x,y)$ which in the 
limit  approaches $\delta(x,y)$,
$$ \widehat{A}(x)^\epsilon\ =\ \int   \text{d}^3y  \widehat{A}(y)\delta^\epsilon(x,y), \ \ \ 
 \widehat{B}(x)^\epsilon\ =\ \int   \text{d}^3y  \widehat{B}(y) \delta^\epsilon(x,y).$$
We obtain  
\begin{align*}\widehat{A}(x)^\epsilon|_{{\cal H}_{\varphi,\{v_1,...,v_m\}}}  \ &=\ 
{\rm id}\otimes\sum_{\alpha=1}^m \delta^\epsilon(x,v_\alpha)\widehat{A}_{v_\alpha}\\
 \widehat{B}(x)^\epsilon|_{{\cal H}_{\varphi,\{v_1,...,v_m\}}} \ &=\ 
{\rm id}\otimes\sum_{\alpha=1}^m \delta^\epsilon(x,v_\alpha)\widehat{B}_{v_\alpha} .
\end{align*}

For sufficiently small $\epsilon$,
$$\widehat{A}(x)^\epsilon\, \widehat{B}(x)^\epsilon\ =\  
{\rm id}\otimes\sum_{\alpha=1}^m \left(\delta^\epsilon(x,v_\alpha)\right)^2\widehat{A}_{v_\alpha}\widehat{B}_{v_\alpha}
$$ 
and on the right hand side at most one term is not zero.  
Since for a quantum representation of a classical $\sqrt{A(x)B(x)}$ we need a symmetric 
operator, we have to apply some symmetric 
product (i.e., symmetrization with respect to different orderings) denoted by ``$:$'' : 
$$:{\widehat{A}}(x)^\epsilon\, \widehat{B}(x)^\epsilon:\ \ =\ 
{\rm id}\otimes\sum_{\alpha=1}^m \left(\delta^\epsilon(x,v_\alpha)\right)^2:\widehat{A}_{v_\alpha}\widehat{B}_{v_\alpha}:.
$$ 
Suppose also that $:\widehat{A}_{v_\alpha}\widehat{B}_{v_\alpha}:$ is self-adjoint. 
Now, we can take the square root (remember: at most one term $\not=0$) and
\begin{align*}
\lim_{\epsilon\rightarrow 0}\ &
\sqrt{:\widehat{A}(x)^\epsilon\, \widehat{B}(x)^\epsilon:}\\
&= \lim_{\epsilon\rightarrow 0} \ {\rm id}\otimes\sum_{\alpha=1}^m \delta^\epsilon(x,v_\alpha)\sqrt{:\widehat{A}_{v_\alpha}\widehat{B}_{v_\alpha}:}\\
&= {\rm id}\otimes\sum_{\alpha=1}^m \delta(x,v_\alpha)\sqrt{:\widehat{A}_{v_\alpha}\widehat{B}_{v_\alpha}:} .
\end{align*} 
\begin{widetext}
Finally, 
\begin{equation*}
\sqrt{u(\widehat{\phi}(x))\otimes {|{\rm det}\widehat{E}(x)|}\ +\  \sqrt{|{\rm det}\widehat{E}(x)|}\widehat{C}_\GR(x)}\ |_{{\cal H}_{\varphi,\{v_1,...,v_m\}}}\ =\ 
{\rm id} \otimes 
\sum_{\alpha=1}^m \delta(x,v_\alpha) \sqrt{: \left(u(\varphi(v_\alpha))\widehat{V}^2_{v_\alpha} + \widehat{C}_{\GR v_\alpha}\right):}
\end{equation*} 

We are now in a position to derive a quantum operator 
\begin{equation}\label{C'''}
\int \text{d}^3x N(x) \sqrt{ - \widehat{\phi}_{,a}\widehat{\phi}_{,b}\widehat{E}^a_{i}\widehat{E}^b_{i} -2u(\widehat{\phi})|\det \widehat{E}| -2 \sqrt{|\det \widehat{E}|}\widehat{C}_\GR}(x).
\end{equation}

We apply the same method of regularization  as in Sec. \ref{A+B},  with $A(x)$ and $B(x)$ of Sec. \ref{A+B} now
replaced by 
\begin{align*} A(x)\ &=\  - {\phi}_{,a}(x){\phi}_{,b}(x){E}^a_{i}(x){E}^b_{i}(x)\\
B(x)\ &=\ -2u({\phi}(x))|\det {E}(x)| -2 \sqrt{|\det {E}(x)|}C_\GR(x), 
\end{align*}
and a state $\ket{\varphi}\otimes\ket{\gamma,j,\iota}$ used in Sec. \ref{A+B} replaced now by  a subspace 
$\eta\left( \ket{\varphi}\otimes {\cal H}_{\gamma} \right)$.   As in Sec. \ref{A+B}, our result takes the following
general form
\begin{align} \label{the result}
&\int \text{d}^3x N(x) \sqrt{ - \widehat{\phi}_{,a}\widehat{\phi}_{,b}\widehat{E}^a_{i}\widehat{E}^b_{i} -2u(\widehat{\phi})|\det \widehat{E}| -2 \sqrt{|\det \widehat{E}|}\widehat{C}_\GR(\widehat{A},\widehat{E}) }(x)\ =\nonumber\\
&\pm i\,\int \text{d}^3x N(x) \sqrt{ \widehat{\phi}_{,a}\widehat{\phi}_{,b}\widehat{E}^a_{i}\widehat{E}^b_{i}}(x) \ +\ 
\int \text{d}^3x N(x) \sqrt{-2u(\widehat{\phi})|\det \widehat{E}| -2 \sqrt{|\det \widehat{E}|}\widehat{C}_\GR}(x).
\end{align} 
The first term on the right hand side is given by (\ref{theoperatordual''}). The second term reads
\begin{align*}
&\int \text{d}^3x N(x) \sqrt{-2u(\widehat{\phi})|\det \widehat{E}| -2 \sqrt{|\det \widehat{E}|}\widehat{C}_\GR}(x)\,|_{{\cal H}_{\varphi,\{v_1,...,v_m\}}}\ \nonumber\\
&=\ \sum_{\alpha=1}^m N(v_\alpha)\sqrt{ -2\, :\left(u(\varphi(v_\alpha))\widehat{V}^2_{v_\alpha} + \widehat{C}_{\GR v_\alpha}\right):}
\end{align*}
 A discussion of the sign of the square rooted expression is in order now.
Classically, the scalar constraint 
$$ C_{\rm tot}(x)\ =\ C_{LQG}(x)\ +\ C_{\rm matt}(x)\ =\ 0$$
implies 
\begin{equation*}
- \phi_{,a}(x)\phi_{,b}(x)E^a_{i}(x)E^b_{i}(x) -2u(\phi(x))|\det E(x)| -2 \sqrt{|\det E(x)|}C_\GR (x) \ \ge\ 0
\end{equation*}
and, because the first term is manifestly non-positive definite, we also have
\begin{equation*}
 -2u(\phi(x))|\det E(x)| -2 \sqrt{|\det E(x)|}C_\GR (x) \ \ge\ 0 .
\end{equation*}
\end{widetext}
After passing to the quantum theory, one could implement those classical inequalities by
assuming that the physical quantum states are defined on the positive part of the spectrum
of the corresponding operators. In our case, however, the quantization took somewhat
unexpected form.  Adapting the above point of view with the given
(\ref{the result}), we have two choices:
\begin{itemize}
\item either admit all the spectrum of the operators\\
\noindent $:\left(u(\varphi(v_\alpha))\widehat{V}^2_{v_\alpha} + \widehat{C}_{\GR v_\alpha}\right):$ at each point $v_\alpha\in\Sigma$
and use it in such a way, that a contribution from the second term on the RHS of  (\ref{the result}) compensates the imaginary terms provided by the first term.    

\item or consider only negative part of the spectrum of the  operators\\
\noindent $:\left(u(\varphi(v_\alpha))\widehat{V}^2_{v_\alpha} + \widehat{C}_{\GR v_\alpha}\right):$ at each point $v_\alpha\in\Sigma$, 
and find a meaning of the factor of $i$ at the first term of (\ref{the result}).
\end{itemize}

The problem in the first choice is that the eigenvalues of the first operator-valued distribution are distributions supported at the edges of $\gamma$ 
while the eigenvalues of the second operator are distributions supported at the vertices. Therefore, to compensate one by 
the other, we would need to introduce a constraint on the lapse functions.  

On the other hand, when making the second choice, we have to take into account the last term of the total constraint, namely  
$$\int N(x)\widehat{\pi}(x) . $$   
This operator is not well defined in the position representation ${\cal H}_\SF$, hence we need to understand better the meaning of 
a quantum constraint which involves it. 
\subsection{Commutators of constraints}
After discussing the status and application in the position representation of the momentum operator
$\widehat{\pi}(x)$ and combining with (\ref{the result}) we spell out the form our quantum constraint operator
$$\widehat{\pi}(x)\ \pm\ \sqrt{ - \widehat{\phi}_{,a}\widehat{\phi}_{,b}\widehat{E}^a_{i}\widehat{E}^b_{i} -2u(\widehat{\phi})|\det \widehat{E}| -2 \sqrt{|\det \widehat{E}|}\widehat{C}_\GR  }(x).  $$
According to (\ref{the result}), the actual structure after quantization is       
\begin{equation} \label{qC}
\begin{split}
\widehat{\pi}(x)\ \pm\Big( 
&\pm i\,\sqrt{ \widehat{\phi}_{,a}\widehat{\phi}_{,b}\widehat{E}^a_{i}\widehat{E}^b_{i}}(x) +\\
&\qquad + 
\sqrt{-2u(\widehat{\phi})|\det \widehat{E}| -2 \sqrt{|\det \widehat{E}|}\widehat{C}_\GR }(x).\Big)
\end{split}
\end{equation} 
where the $\pm$'s come from taking square roots. The clue coming from this form is to study properties of
the operators
\begin{equation*}
\widehat{\pi}\ \pm\  i\,\sqrt{ \widehat{\phi}_{,a}\widehat{\phi}_{,b}\widehat{E}^a_{i}\widehat{E}^b_{i}}(x) 
\end{equation*}
acting by duality on states  $\gbra{\Psi}\otimes \bra{\gamma,j,\iota}\equiv
\sum_{\varphi}\Psi[\varphi]\bra{\varphi}\otimes \bra{\gamma,j,\iota}$. 
We will use the standard notation
\begin{equation*}
\widehat{\pi}^*(M)\equiv \int \text{d}^3x\,M(x)\widehat{\pi}(x)^*
\end{equation*}
etc. Then the basic commutator is
\begin{widetext}
\begin{equation*}
\comm{\widehat{\pi}^*(M)}{\sqrt{\widehat{\phi}_{,a}\widehat{\phi}_{,b}\widehat{E}^a_{i}\widehat{E}^b_{i}}^* (N)}
\gbra{\Psi}\otimes\bra{\gamma,j,\iota}\  =\ \gbra{\Psi'}\otimes\bra{\gamma,j,\iota}
\end{equation*}
where the new functional is given by 
\begin{equation*}
\Psi'[\varphi] = 8\pi \beta \ell_\text{P}^2 \sum_{I}\sqrt{j_I(j_I+1)} \int_{e_I} \text{sign}(d\varphi) N\,\text{d}M\;\Psi[\varphi]
\end{equation*}
Therefore 
\begin{equation*}
\comm{\left(\widehat{\pi}+i\sqrt{ \widehat{\phi}_{,a}\widehat{\phi}_{,b}\widehat{E}^a_{i}\widehat{E}^b_{i}}\right)^*(M)}
{\left(\widehat{\pi}+i\sqrt{ \widehat{\phi}_{,a}\widehat{\phi}_{,b}\widehat{E}^a_{i}\widehat{E}^b_{i}}\right)^*(N)} 
\gbra{\Psi} \otimes\bra{\gamma,j,\iota}\  = \   \gbra{\Psi''} \otimes\bra{\gamma,j,\iota}
\end{equation*}
with
\begin{equation}
\label{eq_comm}
\Psi''[\varphi]=8\pi \beta \ell_\text{P}^2\left(\sum_{I}\sqrt{j_I(j_I+1)} \int_{e_I}\text{sign} (d\varphi) \left(N\,\text{d}M-M\,\text{d}N\right)\right)\Psi[\varphi]
\end{equation}
To calculate commutators of the full constraint, the operators need to act on the spaces ${\cal H}_{\varphi,\{v_1,\ldots,v_m\} }$ to accommodate the gravitational part of the constraint. Here, the function $\varphi$ is adapted to $\{v_1,\ldots,v_m\}$, it has to be constant in a neighborhood of the vertex set. At the same time, we need to sum over all $\varphi$ to have a well defined action of $\widehat{\pi}(x)$. We can achieve both if we additionally assume that the functional $\Psi[\cdot]$ has support only on the functions $\varphi$ that are adapted to the vertex set of a fixed graph in the above sense. 

For the sake of consistency, the operator $\widehat {\pi}(M)$ should then only be applied with $M$ constant in a neighborhood of 
the vertices of the given graph $\gamma$. Then the commutators pass to the states 
$\eta(\bra{\varphi}\otimes\bra{\gamma,j,\iota})$, and ultimately to states of the form
\begin{equation*}
\eta(\gbra{\Psi}\otimes\bra{\gamma,j,\iota}):=\sum_{\varphi} \Psi[\varphi]\eta(\bra{\varphi}\otimes\bra{\gamma,j,\iota}).
\end{equation*}
At this point, we can consider further commutators on these states:           
\begin{equation*}
\comm{\left(\widehat{\pi}+i\sqrt{ \widehat{\phi}_{,a}\widehat{\phi}_{,b}\widehat{E}^a_{i}\widehat{E}^b_{i}}\right)(M)}{ \sqrt{\sqrt{-2|{\rm det}E}|\widehat{C}_\GR}(N)}=0,
\end{equation*}
because $\widehat{\pi}$ and $\sqrt{-2|{\rm det}E}|\widehat{C}_\GR$ act on different tensor factors in the Hilbert space, and $\sqrt{ \widehat{\phi}_{,a}\widehat{\phi}_{,b}\widehat{E}^a_{i}\widehat{E}^b_{i}})$ does not evaluate or change the structure at the vertices, whereas the gravitational part of the constraint does so exclusively. Moreover
\begin{equation*}
\comm{\sqrt{\sqrt{-2|{\rm det}E}|\widehat{C}_\GR} (M)}{\sqrt{\sqrt{-2|{\rm det}E}|\widehat{C}_\GR}(N)}=0
\end{equation*}
is ensured by suitable construction of $\widehat{C}_\GR$ \cite{Thiemann:1996aw,Thiemann:1996av,Thiemann:1997rv,Lewandowski:1997ba,arXiv:1410.5276}. Finally, in the case of a nonzero potential $u(\varphi)$, 
and for a gravitational state $\ket{\gamma, \{V_v\}}$ that is an eigenstate of volume, 
\begin{equation*}
\sqrt{|\det\widehat{E}| }(M)\ket{\gamma, \{V_v\}}=\sum_{v\in V(\gamma)} M(v)V_v \ket{\gamma, \{V_v\}}
\end{equation*}
we find
\begin{equation*}
\comm{\pi(M)}{\sqrt{u(\widehat{\phi}(x))\otimes {|\det\widehat{E}|}}(M)}\,
\gbra{\Psi}\otimes\bra{\gamma , \{V_v\}}
=\gbra{\Psi'''}\otimes\bra{\gamma , \{V_v\}}
\end{equation*}
with 
\begin{equation*}
\Psi'''[\varphi]=\sum_v M(v)N(v)\, \sqrt{u'(\varphi(v))} V_v, 
\end{equation*}
i.e., the commutator is symmetric in $M$ and $N$.  
Consider now the total quantum scalar constraint 
\begin{equation*}
\widehat{C}_{\rm tot}(M)\ =\ \int \text{d}^3x M(x)\left(\widehat{\pi}(x)\ +  i\,\sqrt{ \widehat{\phi}_{,a}\widehat{\phi}_{,b}\widehat{E}^a_{i}\widehat{E}^b_{i}}(x) \ +\ 
\sqrt{-2u(\widehat{\phi})|\det \widehat{E}|-2 \sqrt{|\det \widehat{E}|}\widehat{C}_\GR }(x)\right) 
\end{equation*} 
Taking together all the commutators that we have obtained before, we have 
\begin{equation}
\label{eq_comm2}
\comm{\widehat{C}_{\rm tot}(M)}{\widehat{C}_{\rm tot}(N)}\eta(\gbra{\Psi}\otimes\bra{\gamma,j,\iota})
=\eta(\gbra{\Psi''}\otimes\bra{\gamma,j,\iota})
\end{equation}
with $\Psi''$ given in \eqref{eq_comm}. This result is a blessing as well as a curse. It is a blessing because the operator defined by
\begin{equation*}
D(M,N)\, \eta(\gbra{\Psi} \otimes\bra{\gamma,j,\iota}) = \eta(\gbra{\Psi''}\otimes\bra{\gamma,j,\iota})
\end{equation*}
can be understood as the quantization of the scalar field part of the diffeomorphism constraint
\begin{equation*}
D_\SF(S)= \int  S^a(x)\;  \pi \partial_a\phi (x)\;  \text{d}^3 x, \qquad S^a = \frac{E_i^a E_j^b\delta^{ij}}{|\det E |} \left(N\partial_bM-M\partial_bN\right) 
\end{equation*}
Indeed, in the standard interpretation of the quantum states $\bra{\gamma,j,\iota}$, the quantized metric has support only on the edges of $\gamma$, which explains the restriction of the integration in \eqref{eq_comm} to edges. There are also puzzling aspects. What is perhaps most troubling is that $D(M,N)$ acts on states that are supposedly invariant under a large class of diffeomorphisms. One would have to show that those generated by $S^a$ above are not among them. This question, the commutators of $D(M,N)$, and further aspects are currently under study. 

\eqref{eq_comm2} is also a curse, because it means that solutions to the quantum constraint $\widehat{C}_{\rm tot}$ can not be found among the states considered so far. Indeed
\begin{equation*}
\widehat{C}_{\rm tot}(M)\eta(\gbra{\Psi}\otimes\bra{\gamma,j,\iota})=0 \quad \Longrightarrow D(M,N)\eta(\gbra{\Psi}\otimes\bra{\gamma,j,\iota})=0. 
\end{equation*}
But the latter implies 
\begin{equation*}
8\pi \beta \ell_\text{P}^2\left(\sum_{I}\sqrt{j_I(j_I+1)} \int_{e_I}\text{sign} (d\varphi) \left(N\,\text{d}M-M\,\text{d}N\right)\right)\Psi[\varphi]\ =\ 0.
\end{equation*}   
Since the lapse functions $N$ and $M$ are arbitrary, this is a very strong condition, one that has no nontrivial solutions among the class of states considered in this article. 
\end{widetext}
\section{Summary and outlook}
We have reconsidered the the Rovelli-Smolin model of gravity coupled to the Klein-Gordon time field with an eye towards capturing the degrees of freedom of the scalar field lost in the framework in which time is deparametrized by the scalar field. 

While we have not fully solved this problem, we have obtained several new results:
\begin{itemize}
\item We have constructed a kinematic Hilbert space for the gravity-matter system and a non-standard diffomorphism invariant representation of the scalar field thereon. Going over to the dual space, the field momentum is well defined along with the field itself. The dual states can be group averaged to be partially diffeomorphism invariant. 
\item The new representation for the scalar field comes with a dual Hilbert space representation which relates to it like the momentum representation relates to the position representation in quantum mechanics. However, in the present case the duality is not a unitary map. The standard representation \cite{Thiemann:1997rq} used for scalars in LQG is a subrepresentation of this dual. 
\item We obtained a new operator for the scalar constraint of the coupled system. Commutators of this operator are well defined and reproduce part of the Dirac algebra.  
\item We have developed new methods for solving the constraint equation. 
\item We have described a representation of the gravitational degrees of freedom in which the flux is diagonal. This representation bears a strong resemblance to the BF vacuum of Dittrich and Geiller.
\end{itemize}
The fact that commutators of the new constraint do not vanish, poses problems for finding solutions to the constraint. Hence the states we consider -- and perhaps the whole setup -- still needs some improvement. Areas that merit further study are the higher order commutators of the constraint, and their relation to the algebra of diffeomorphism generators, findings solutions to the full constraint, and ultimately the connection to the quantum theory of the deparametrized version of the system. 

\section*{Acknowledgements}
This work was partially supported by the grant of the Polish Narodowe Centrum Nauki nr
2011/02/A/ST2/00300, Foundation for Polish Science and by the Emerging Fields Project \emph{Quantum Geometry} of the Friedrich-Alexander University Erlangen-N\"urnberg. JL thanks members of Institute for Quantum Gravity at the Friedrich-Alexander University Erlangen-N\"urnberg where this work was started for hospitality. HS would also like to thank organizers of the 1st Conference of the Polish Society on GR, where part of this work was completed, for hospitality.


\begin{thebibliography}{12}

\bibitem{gr-qc/9308002}
 C.~Rovelli and L.~Smolin,
 ``The Physical Hamiltonian in nonperturbative quantum gravity,''
 Phys.\ Rev.\ Lett.\  {\bf 72}, 446 (1994)
 [arXiv:gr-qc/9308002].

\bibitem{arXiv:1009.2445}
 M.~Domagala, K.~Giesel, W.~Kaminski and J.~Lewandowski,
 ``Gravity quantized: Loop Quantum Gravity with a Scalar Field,''
 Phys.\ Rev.\  D {\bf 82}, 104038 (2010)
 [arXiv:1009.2445 [gr-qc]].

\bibitem{gr-qc/9409001}
 J.~D.~Brown and K.~V.~Kuchar,
 ``Dust as a standard of space and time in canonical quantum gravity,''
 Phys.\ Rev.\  D {\bf 51}, 5600 (1995)
 [arXiv:gr-qc/9409001].

\bibitem{arXiv:0711.0119}
 K.~Giesel and T.~Thiemann,
 ``Algebraic quantum gravity (AQG). IV. Reduced phase space quantisation of
 loop quantum gravity,''
 Class.\ Quant.\ Grav.\  {\bf 27}, 175009 (2010)
 [arXiv:0711.0119 [gr-qc]].

\bibitem{arXiv:1108.1145}
 V.~Husain and T.~Pawlowski,
 ``Time and a physical Hamiltonian for quantum gravity,''
 Phys.\ Rev.\ Lett.\  {\bf 108}, 141301 (2012)
 [arXiv:1108.1145 [gr-qc]].

\bibitem{arXiv:0906.2798}
 G.~M.~Hossain, V.~Husain and S.~S.~Seahra,
 ``Non-singular inflationary universe from polymer matter,''
 Phys.\ Rev.\  D {\bf 81}, 024005 (2010)
 [arXiv:0906.2798 [astro-ph.CO]].

\bibitem{arXiv:1002.0138}
 V.~Husain and A.~Kreienbuehl,
 ``Ultraviolet behavior in background independent quantum field theory,''
 Phys.\ Rev.\  D {\bf 81}, 084043 (2010)
 [arXiv:1002.0138 [gr-qc]].

\bibitem{arXiv:1306.6126}
 M.~Varadarajan,
 ``The generator of spatial diffeomorphisms in the Koslowski- Sahlmann
 representation,''
 Class.\ Quant.\ Grav.\  {\bf 30}, 175017 (2013)
 [arXiv:1306.6126 [gr-qc]].

\bibitem{arXiv:1311.6117}
 M.~Campiglia and M.~Varadarajan,
 ``The Koslowski- Sahlmann representation: Gauge and diffeomorphism
 invariance,''
 Class.\ Quant.\ Grav.\  {\bf 31}, 075002 (2014)
 [arXiv:1311.6117 [gr-qc]].

\bibitem{arXiv:1406.0579}
 M.~Campiglia and M.~Varadarajan,
 ``The Koslowski-Sahlmann representation: Quantum Configuration Space,''
 Class.\ Quant.\ Grav.\  {\bf 31}, 175009 (2014)
 [arXiv:1406.0579 [gr-qc]].

\bibitem{Ashtekar:2002vh}
 A.~Ashtekar, J.~Lewandowski and H.~Sahlmann,
 ``Polymer and Fock representations for a scalar field,''
 Class.\ Quant.\ Grav.\  {\bf 20}, L11 (2003)
 [arXiv:gr-qc/0211012].

\bibitem{gr-qc/0005117}
 Y.~Ma and Y.~Ling,
 ``The Q-hat operator for canonical quantum gravity,''
 Phys.\ Rev.\  D {\bf 62}, 104021 (2000)
 [arXiv:gr-qc/0005117].

\bibitem{arXiv:1401.6441}
 B.~Dittrich and M.~Geiller,
 ``A new vacuum for Loop Quantum Gravity,''
 arXiv:1401.6441 [gr-qc].

\bibitem{arXiv:1412.3752}
 B.~Dittrich and M.~Geiller,
 ``Flux formulation of loop quantum gravity: Classical framework,''
 arXiv:1412.3752 [gr-qc].

\bibitem{gr-qc/0404018}
 A.~Ashtekar and J.~Lewandowski,
 ``Background independent quantum gravity: A Status report,''
 Class.\ Quant.\ Grav.\  {\bf 21}, R53 (2004)
 [arXiv:gr-qc/0404018].

\bibitem{arXiv:0904.0184}
 A.~Ashtekar,
 ``Some surprising implications of background independence in canonical
 quantum gravity,''
 Gen.\ Rel.\ Grav.\  {\bf 41}, 1927 (2009)
 [arXiv:0904.0184 [gr-qc]].

\bibitem{Koslowski:2007kh}
 T.~A.~Koslowski,
``Dynamical Quantum Geometry (DQG Programme),''
 arXiv:0709.3465 [gr-qc].

\bibitem{Sahlmann:2010hn}
 H.~Sahlmann,
 ``On loop quantum gravity kinematics with non-degenerate spatial
 background,''
 Class.\ Quant.\ Grav.\  {\bf 27}, 225007 (2010)
 [arXiv:1006.0388 [gr-qc]].

\bibitem{arXiv:1312.3657}
 A.~Stottmeister and T.~Thiemann,
 ``Structural aspects of loop quantum gravity and loop quantum cosmology from
 an algebraic perspective,''
 arXiv:1312.3657 [gr-qc].

\bibitem{Ashtekar:1994wa}
 A.~Ashtekar and J.~Lewandowski,
 ``Differential geometry on the space of connections via graphs and projective
 limits,''
 J.\ Geom.\ Phys.\  {\bf 17}, 191 (1995)
 [arXiv:hep-th/9412073].

\bibitem{Lewandowski:2005jk}
 J.~Lewandowski, A.~Okolow, H.~Sahlmann and T.~Thiemann,
 ``Uniqueness of diffeomorphism invariant states on holonomy-flux algebras,''
 Commun.\ Math.\ Phys.\  {\bf 267}, 703 (2006)
 [arXiv:gr-qc/0504147].

\bibitem{Ashtekar:1994mh}
 A.~Ashtekar and J.~Lewandowski,
 ``Projective techniques and functional integration for gauge theories,''
 J.\ Math.\ Phys.\  {\bf 36}, 2170 (1995)
 [arXiv:gr-qc/9411046].

\bibitem{Rovelli:1989za}
 C.~Rovelli and L.~Smolin,
 ``Loop Space Representation of Quantum General Relativity,''
 Nucl.\ Phys.\  B {\bf 331}, 80 (1990).

\bibitem{Thiemann:2000bv}
 T.~Thiemann,
 ``Quantum spin dynamics (QSD): 7. Symplectic structures and continuum lattice
 formulations of gauge field theories,''
 Class.\ Quant.\ Grav.\  {\bf 18}, 3293 (2001)
 [arXiv:hep-th/0005232].

\bibitem{arXiv:1410.5276}
 J.~Lewandowski and H.~Sahlmann,
 ``A symmetric scalar constraint for loop quantum gravity,''
 arXiv:1410.5276 [gr-qc].

\bibitem{Thiemann:1996aw}
 T.~Thiemann,
 ``Quantum spin dynamics (QSD),''
 Class.\ Quant.\ Grav.\  {\bf 15}, 839 (1998)
 [arXiv:gr-qc/9606089].

\bibitem{Thiemann:1996av}
 T.~Thiemann,
 ``Quantum spin dynamics (qsd). 2,''
 Class.\ Quant.\ Grav.\  {\bf 15}, 875 (1998)
 [arXiv:gr-qc/9606090].

\bibitem{Thiemann:1997rv}
 T.~Thiemann,
 ``QSD 3: Quantum constraint algebra and physical scalar product in quantum
 Class.\ Quant.\ Grav.\  {\bf 15}, 1207 (1998)
 [arXiv:gr-qc/9705017].

\bibitem{Lewandowski:1997ba}
 J.~Lewandowski and D.~Marolf,
 ``Loop constraints: A Habitat and their algebra,''
 Int.\ J.\ Mod.\ Phys.\  D {\bf 7}, 299 (1998)
 [arXiv:gr-qc/9710016].
 
\bibitem{Thiemann:1997rq}
  T.~Thiemann,
  ``Kinematical Hilbert spaces for Fermionic and Higgs quantum field theories,''
  Class.\ Quant.\ Grav.\  {\bf 15} (1998) 1487
  [gr-qc/9705021].
 
 \end{thebibliography}
\end{document}